\documentclass[prd,preprint,superscriptaddress,showpacs,nofootinbib,%
tightenlines]{revtex4}
\usepackage{mathrsfs}
\usepackage{latexsym,bm}
\usepackage{CJK}
\usepackage{graphicx}
\usepackage{indentfirst}
\usepackage{slashed}
\usepackage{amsmath}
\usepackage{amssymb}
\usepackage{color}
\usepackage{hyperref}
\usepackage{epsfig}
\hypersetup{CJKbookmarks=true}
\usepackage[titletoc]{appendix}
\usepackage{multirow}%
\usepackage{rotating}
\usepackage{epstopdf}
\usepackage{extarrows}
\usepackage{tabularx}

\renewcommand{\arraystretch}{1.5}

\newcommand{\be}{\begin{equation}} \newcommand{\ee}{\end{equation}}
\newcommand{\ba}{\begin{array}{c}} \newcommand{\ea}{\end{array}}
\newcommand{\bea}{\begin{eqnarray}} \newcommand{\eea}{\end{eqnarray}}

\newcommand{\rd}{{\rm d}}
\newcommand{\order}[1]{\mathcal{O}\left(#1\right)}
\newcommand{\al}{&\!\!\!\!}

\newcommand{\Lag}{\mathcal{L}}

\begin{document}
\title{
Aspects of the low-energy constants in the chiral
Lagrangian for charmed mesons}
\author{Meng-Lin Du}
\email{du@hiskp.uni-bonn.de}
\affiliation{Helmholtz-Institut f\"ur Strahlen- und Kernphysik and
Bethe Center for Theoretical Physics, Universit\"at~Bonn, D--53115
Bonn, Germany}
\author{Feng-Kun Guo}
\email{fkguo@itp.ac.cn}
\affiliation{CAS Key Laboratory of Theoretical Physics, Institute of
Theoretical Physics, Chinese~Academy~of~Sciences,
Zhong~Guan~Cun~East~Street~55, Beijing~100190, China}
\author{Ulf-G. Mei{\ss}ner}
\email{meissner@hiskp.uni-bonn.de}
\affiliation{Helmholtz-Institut f\"ur Strahlen- und Kernphysik and
Bethe Center for Theoretical Physics, Universit\"at~Bonn, D--53115
Bonn, Germany}
\affiliation{Institute for Advanced Simulation, Institut f{\"u}r
Kernphysik and J\"ulich Center for Hadron Physics, Forschungszentrum~
J{\"u}lich, D-52425 J{\"u}lich, Germany}
\author{ De-Liang Yao}
\email{d.yao@fz-juelich.de}
\affiliation{Institute for Advanced Simulation, Institut f{\"u}r
Kernphysik and J\"ulich Center for Hadron Physics, Forschungszentrum~
J{\"u}lich, D-52425 J{\"u}lich, Germany}

\date{\today}

\medskip

\begin{abstract}

We investigate the numerical values of the low-energy constants in the chiral
effective Lagrangian for the interactions between the charmed
mesons and the lightest pseudoscalar mesons,  the Goldstone bosons of the
spontaneous breaking of chiral symmetry for QCD. This problem is tackled from two
sides:
estimates using the resonance exchange model, and positivity constraints from
the general properties of the $S$-matrix including analyticity, crossing
symmetry and unitarity.
These estimates and constraints are compared with the
values determined from fits to  lattice data of the scattering lengths.
Tensions are found, and possible reasons are discussed. We conclude that
more data from lattice calculations and experiments are necessary to fix these
constants better. As a by-product, we also estimate the coupling constant
$g_{DDa_2}$, with $a_2$ the light tensor meson, via the QCD sum rule approach.

\end{abstract}

\pacs{12.39.Fe, 13.75.Lb, 14.40.Lb}

\maketitle

\section{Introduction}
\label{sec:intro}

Chiral perturbation theory
(ChPT)~\cite{Weinberg:1978kz,Gasser:1983yg,Gasser:1984gg}, the low-energy
effective field theory (EFT) of quantum chromodynamics (QCD), nowadays plays a
crucial role in studying hadron physics at low energies.
It is based on the spontaneous breaking of the  approximate QCD chiral symmetry
SU$(N_f)_L\times$SU$(N_f)_R$, where $N_f$ is the number of light flavors, down to its
vectorial subgroup SU$(N_f)_V$. The lightest pseudoscalar mesons, much lighter
than any other hadron, appear as the Goldstone bosons which are the effective
degrees of freedom of ChPT.
As a typical EFT, it accounts for  the separation of energy scales in the
physical systems under consideration:
only the low-energy Goldstone modes are treated explicitly (external sources can
be included easily), while the information of any other QCD excitation (at
scales $\gtrsim\Lambda_\chi\sim1$~GeV) is encoded in the coefficients in front
of the local operators constructed from the Goldstone fields, which are unknown
parameters and called low-energy constants (LECs) in ChPT.
The  determination of the chiral LECs is an important issue because it is
essential for the predictive power of ChPT and further can serve as a 
consistency check of the theory.

Ideally, the LECs should be pinned down by comparing with (or performing fits
to) experimental or lattice QCD data of selected observables in certain
processes. Since these values of LECs should be universal, consequently,
predictions for other processes or physical quantities can be made.
For instance, the LECs of the fundamental $\pi N$
interaction~\cite{Fettes:1998ud,Fettes:2000xg,Chen:2012nx,Alarcon:2012kn,Hoferichter:2015tha,Yao:2016vbz},
which are fixed by fitting to experimental $\pi N$ scattering data, are employed
to make predictions in $\pi\pi N$ physics, see e.g.Ref.~\cite{Fettes:1999wp} and $NN$
physics, see Ref.~\cite{Epelbaum:2008ga} for a review.
However, things become cumbersome when there are not sufficiently many data or,
even worse, no good data for fixing the LECs. Furthermore, even if the LECs have
been extracted or estimated using some procedure, the reliability of these
values still needs to be further analyzed.

A phenomenological approach to estimate the LECs was discussed in detail in
Refs.~\cite{Ecker:1988te,Ecker:1989yg,Donoghue:1988ed}, and is traditionally
referred to as the resonance saturation.
Therein, phenomenological Lagrangians respecting chiral symmetry including
explicit meson resonances are constructed and then the resonance fields are
integrated out to generate contributions to the LECs in the mesonic ChPT
Lagrangian at tree level in terms of the resonance couplings and masses.  It was
found that whenever the vector and axial vector mesons contribute,  they almost
saturate the empirical values of the LECs, which is a modern version of the
vector meson dominance hypothesis.
Similarly, the resonance exchange model  also provides a fairly good
phenomenological description of the LECs in the chiral Lagrangian for the
pion-nucleon interactions, c.f.~Ref.~\cite{Bernard:1996gq}, where it is found
that the $\Delta$ resonance provides the dominant contribution to some of the
LECs, i.e., $c_3$ and $c_4$.\footnote{Note that there is also an important
contribution from the $\rho$-meson to $c_4$.} In view of the success achieved
in both the purely mesonic ChPT and baryon ChPT, the resonance exchange method
will be discussed in this paper to estimate the LECs related to the interactions
between charmed $D$ mesons and Goldstone bosons (to be denoted as $\phi$), which
are badly known because no experimental data for t $D$$\phi$ scattering are
available and almost all the existing extractions result from fitting to lattice
results of scattering lengths for certain channels.

The $D$$\phi$ interaction is of great importance in understanding the
heavy-light meson spectrum on the one hand, and serves as an ideal playground to
combine heavy quark symmetry and chiral symmetry on the other one.
A good example for the former is provided by the $D_{s0}^\ast(2317)$ discovered
in 2003~\cite{Aubert:2003fg,Krokovny:2003zq}. As this state might be
a $DK$ bound state~\cite{Barnes:2003dj}, progress towards unravelling its nature
has been made along the line of studying the interaction between the $D$ meson
and the
kaon~\cite{Kolomeitsev:2003ac,Hofmann:2003je,Guo:2006fu,Gamermann:2006nm,
Guo:2009ct,Guo:2008gp,Liu:2012zya,Mohler:2013rwa,Moir:2016srx}.

The $D$$\phi$ interaction can also give guidance for $D^\ast$$\phi$, $B$$\phi$
and $B^\ast$$\phi$ interactions, since similarities amongst them exist due to
heavy quark spin and flavor
symmetries~\cite{Burdman:1992gh,Wise:1992hn,Yan:1992gz}.
The heavy quark symmetries relate the pseudoscalar $D$ mesons to the vector
$D^*$ as well as to the bottom analogues. Thus, the LECs determined in one
sector can be used in the other heavy-quark-symmetry-related sectors at leading
order of the heavy quark expansion once the heavy quark mass scaling is properly taken into account.

Partly stimulated by the lattice QCD results for the $D\phi$ scattering lengths
in the past few
years~\cite{Liu:2008rza,Liu:2012zya,Mohler:2012na,Mohler:2013rwa}, the $D\phi$
interaction has been revisited using the chiral Lagrangian up to the
next-to-leading order (NLO) or the next-to-next-to-leading order (NNLO), and the 
LECs are determined from fitting to the lattice results of the scattering
lengths using either perturbative~\cite{Liu:2009uz,Geng:2010vw} or unitarized
scattering
amplitudes~\cite{Liu:2012zya,Wang:2012bu,Altenbuchinger:2013vwa,Yao:2015qia,
Guo:2015dha}.
However, the scarcity of data and the model dependence of the unitarization
method cause discrepancies among the extracted values for some of the LECs.

Furthermore, we will investigate model-independent positivity constraints on the
$D$$\phi$ interaction as well. Similar to the case for the
$\pi\pi$~\cite{Ananthanarayan:1994hf,Dita:1998mh,Manohar:2008tc,Mateu:2008gv,
Guo:2009vf}
and $\pi N$~\cite{Luo:2006yg,Sanz-Cillero:2013ipa} scattering, these constraints
will be derived in the  upper part of the Mandelstam triangle (with $t>0$)
based on axiomatic principles of the $S$-matrix theory such as analyticity,
unitarity and crossing symmetry. After applying the obtained constraints to the
chiral perturbative and EOMS-renormalized amplitudes, e.g., given by
Ref.~\cite{Yao:2015qia}, one obtains restrictions on the involved LECs at a
certain given order. These axiomatic constraints will be confronted with the
numerical values of the LECs determined through various phenomenological fits to
lattice data.

This paper is structured as follows. In Section~\ref{sec:DphiS}, we start with a
brief review of the formal aspects of $D$$\phi$ scattering. In
Section~\ref{sec:resonance}, we discuss the possible resonances that should be
taken into  consideration and introduce the relevant phenomenological
Lagrangians for resonance exchange. Then, in Section~\ref{sec:contribution}, we
compute the resonance contributions to the LECs at tree level analytically and
numerically. In Section~\ref{SecPos}, we deal with the axiomatic constraints on
the $D\phi$ scattering amplitudes in perturbation theory, which are finally
transformed into positivity bounds on the LECs, and we will compare these
constraints with the determinations in nonperturbative fits to lattice data.
A summary of this work is given in Section~\ref{SecCon}. The leading order (LO)
Born-term contributions to $D\phi$ scattering are relegated to
Appendix~\ref{app:LOBorn} for completeness.
We collect the contribution to LECs from the exchange of light tensor mesons and
an estimate of their coupling to the $D$-meson $g_{DDT}$ using the  QCD sum rule
approach in the Appendices~\ref{app:h5} and \ref{app:sumrule}, respectively.

\section{Low-energy constants and resonance exchanges
\label{sec:ResEx}}

In this section, we give a short introduction to some relevant issues related to
$D$$\phi$ scattering, such as the involved LECs and the related Mandelstam
plane. Then the LO chiral Lagrangians for various resonances are discussed
and constructed for later use. Finally, we make use of the approach of resonance
exchange to analyze the resonance contributions to the LECs. The corresponding
numerical results are also given.

\subsection{{\boldmath$D\phi$} scattering at low energies
\label{sec:DphiS}}

As mentioned in the Introduction, the pseudoscalar and vector
charmed mesons are related to each other via heavy quark spin symmetry. One can
construct ChPT for heavy mesons by treating the pseudoscalars and the vectors
simultaneously in a spin multiplet. The scattering processes of the Goldstone
bosons off the pseudoscalar and vector charmed mesons can thus be described by
the same chiral Lagrangian at low energies. The LECs of  such a Lagrangian
will be discussed in this paper. So far, most of the available information
for these LECs was obtained from fitting to the lattice data of the $S$-wave
scattering lengths for the $D\phi$ systems~\cite{Liu:2012zya,Yao:2015qia}. It
has been shown by explicit calculations that the $D^*$ contribution is
negligible in these quantities~\cite{Cleven:2010aw,Yao:2015qia}. Hence, we can
focus on the Lagrangian without the $D^*$, and keep in mind that such a theory
is basically equivalent to the one with the $D^*$ explicitly included when
discussing the $S$-wave $D\phi$ scattering.
The relevant chiral effective Lagrangian
reads~\cite{Liu:2012zya,Yao:2015qia},\footnote{There are in fact four more terms
at $\order{p^3}$ which can be found in Ref.~\cite{Du:2016ntw}. However, they do
not contribute to the $D\phi$ scattering, and thus will not be discussed here.}
\bea
\mathcal{L}_{D\phi} \al=\al \mathcal{D}_\mu D \mathcal{D}^\mu D^\dagger-
\overset{\circ\;}{M^2}D D^\dagger\nonumber\\
\al+\al D\left(-h_0\langle\chi_+\rangle-h_1{\chi}_+
+ h_2\langle u_\mu u^\mu\rangle-h_3u_\mu u^\mu\right) {D}^\dag
 + \mathcal{D}_\mu D\left({h_4}\langle u_\mu
u^\nu\rangle-{h_5}\{u^\mu,u^\nu\}\right)\mathcal{D}_\nu {D}^\dag\ \nonumber \\
\al + \al
D\biggl[ i\,{g_1}[{\chi}_-,u_\nu] +
{g_2}\left([u_\mu,[\mathcal{D}_\nu,u^\mu]] + [u_\mu,[\mathcal{D}^\mu,u_\nu]]
\right)\biggr]\mathcal{D}^\nu {D}^\dag + g_3
D\,[u_\mu,[\mathcal{D}_\nu,u_\rho]] \mathcal{D}^{\mu\nu\rho} {D}^\dag
\nonumber\\
\al+\al \text{higher-order terms}, \label{eq:LDphi}
\eea
where the building blocks are given by
\bea
u_\mu = i\left(u^\dagger\partial_\mu u-u\,\partial_\mu u^\dagger\right)\ ,
\quad
\chi_\pm = u^\dagger\chi u^\dagger\pm u\chi^\dagger u\ ,
\eea
with $\chi=2B_0\,{\rm diag}(m_u,m_d, m_s)$ and
\bea
u=\exp\left(\frac{i\phi}{\sqrt{2}F_0}\right)\ , \quad
\phi=\begin{pmatrix}
   \frac{1}{\sqrt{2}}\pi^0 +\frac{1}{\sqrt{6}}\eta  & \pi^+ &K^+  \\
     \pi^- &  -\frac{1}{\sqrt{2}}\pi^0 +\frac{1}{\sqrt{6}}\eta&K^0 \\
     K^-&\bar{K}^0&-\frac{2}{\sqrt{6}}\eta
\end{pmatrix} .
\eea
Here, $B_0$ is a constant related to the quark condensate. $F_0$ and
$\overset{\circ}{M}$ are the pion decay constant and the mass of pseudoscalar
charmed mesons  in the chiral limit, respectively. The coupling constants
$h_i,~g_i$ are the  LECs to be discussed in this paper.  $D$ denotes
the SU(3) triplet of the ground state pseudoscalar charmed mesons, i.e.,
$D=(D^0,D^+,D_s^+)$.
${\cal D}_\mu$ is the chirally covariant derivative acting on the $D$-meson
fields and $\mathcal{D}^{\mu\nu\rho}=\{\mathcal{D}_\mu, \{\mathcal{D}_\nu,
\mathcal{D}_\rho\}\}$.

It is worth noting that the mass dimensions of the LECs $h_{i=0,\cdots,5}$ as
defined in Refs.~\cite{Liu:2012zya,Yao:2015qia} are different, so are those for
$g_{j=1,2,3}$. Therefore, in Refs.~\cite{Liu:2012zya,Yao:2015qia}, the following
redefinitions of the LECs are employed:
\bea\label{eq:LECsCom}
h_{4,5}^\prime \al=\al  h_{4,5}\, \bar{M}_D^2\ ,\quad
h_{24} = h_2+h_4^\prime\ ,\quad h_{35}=h_3+2h_5^\prime\ ,
\nonumber\\
g_{23}\al=\al g_2^\prime-2g_3^\prime\ ,\quad g_{1,2}^\prime=g_{1,2}\, \bar{M}_D\
,\quad g_3^\prime=g_3 \bar{M}_D^3\ ,
\eea
where $\bar{M}_D=(M_D^{\rm
phy}+M_{D_s}^{\rm phy})/2$, with $M_D^{\rm phy}$ and $M_{D_s}^{\rm phy}$ the
physical masses of the $D$ and $D_s$ mesons, respectively.
These newly defined coefficients of the ${\cal O}(p^2)$ and
${\cal O}(p^3) $ operators are in units of $1$ and ${\rm GeV}^{-1}$,
respectively. The  $h_{0,1}$ are dimensionless, too. As discussed in
Ref.~\cite{Liu:2012zya}, such redefinitions are also designed to reduce the
correlations between the LECs, and are useful and necessary to obtain reliable
numeral results when performing fits.

For  $D\phi$ scattering, there are in total 16 channels with different
strangeness $S$ and isospin $I$ quantum numbers. Nevertheless, since the scattering amplitudes
${\cal A}(s,t)$ are related to each other according by crossing
symmetry, only 10 amplitudes
are independent in the end and can be taken as the basis to construct the other
amplitudes. The 10 amplitudes calculated in the physical particle bases
following Refs.~\cite{Guo:2009ct,Yao:2015qia} are given in
Appendix~\ref{app:LOBorn}. They will be used when deriving the
resonance-exchange amplitudes in Section~\ref{sec:resonance}.

\begin{figure}[t]
\begin{center}
\epsfig{file=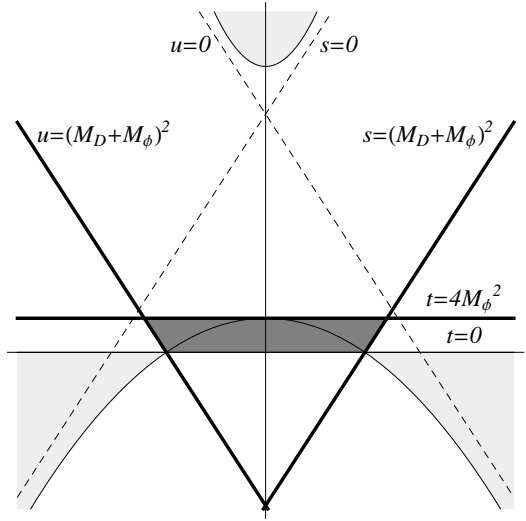,scale=0.65}
\end{center}
\caption{The Mandelstam plane.
The Mandelstam triangle is the region bounded by the thick lines: $s=(M_D+M_\phi)^2$,
$u=(M_D+M_\phi)^2$ and $t=4M_\phi^2$. The upper part of the Mandelstam triangle is marked
in dark gray, which is surrounded by the previous three lines and the one corresponding to  $t=0$. The physical regions
are marked in light gray.}\label{fig:Mandelstam}
\end{figure}

Before ending this subsection, let us introduce the Mandelstam plane which will
be used when deriving the positivity constraints. In the Mandelstam $s$-$t$
plane, the kinematical region of the $D\phi$ scattering is defined as the domain
where the Kibble function~\cite{Kibble:1960zz}
$\Phi=t\left[su-(M_D^2-M_\phi^2)^2\right]$ is non-negative. The plane is
depicted in Fig.~\ref{fig:Mandelstam}, where the bottom-right, bottom-left and
top areas in light gray denote the $s$-, $u$- and $t$-channel physical regions,
respectively. The interior of the triangle surrounded by lines of
$s=(M_D+M_\phi)^2$, $u=(M_D+M_\phi)^2$ and $t=4M_\phi^2$ is called the
Mandelstam triangle, where the scattering amplitude is real and analytical.
The positivity constraints will be calculated in the upper part of the
Mandelstam triangle where $t\geq 0$, see the area marked in dark gray in
Fig.~\ref{fig:Mandelstam}.  Notice that the condition $t\geq 0$ guarantees that
the Legendre polynomials $P_\ell(\cos\theta)$, with $\theta$ the scattering
angle, are non-negative, which is necessary for deriving the positivity
constraints~\cite{Manohar:2008tc,Mateu:2008gv,Sanz-Cillero:2013ipa}.

\subsection{Chiral resonance Lagrangians\label{sec:resonance}}

The saturation of LECs by the contributions from resonances is based on scale
separation such that the low-energy effective Lagrangian contains only the
low-lying degrees of freedom and the resonances at the hard scale are considered
to be integrated out. The local operators in the Lagrangian are constructed
in terms of the effective degrees of freedom, while the high-energy contribution
including the effects from resonances enter the LECs, which are coefficients of
the operators. In principle, the LECs can be calculated in the full theory by a
matching procedure. In the case of the chiral Lagrangian, since we cannot solve
the nonperturbative QCD analytically, we may match the chiral Lagrangian
containing only the low-lying degrees of freedom to the one with resonances,
which is applicable in a larger energy range phenomenologically despite the more
complicated renormalization and power counting issues related to the large
masses and instability of the resonances.
For such a matching, one expects that the resonances with relatively low masses
contribute dominantly to the LECs.

To analyze the resonance contributions to the chiral LECs, the chiral resonance
Lagrangians are necessary.  We will first introduce the Lagrangians related to
excited charmed mesons, with the  orbital angular momentum between the charmed
quark and light quark $\ell\leq1$, then the ones concerning the light-flavor
mesonic excitations will be discussed.

The excited charmed mesons with $\ell\leq1$ include $D_0^\ast$ with $J^P=0^+$,
$D_1^\prime$ and $D_1$ with $J^P=1^+$, and $D_2^\ast$ with
$J^P=2^+$.\footnote{Note that the $D^\ast$ vector mesons with $J^P=1^-$ are
treated as the spin partner of the $D$ as discussed at the beginning of this
section.} Though more and more candidates for states with $\ell\geq 2$ were
discovered experimentally~\cite{Aubert:2006mh,Brodzicka:2007aa,Aubert:2009ah},
their classifications in the charmed spectra still need to be investigated or
confirmed. Furthermore, their contributions should be smaller than those with
$\ell\leq1$ because of higher masses as mentioned above. Hence we do not include
them in our analysis.

For the scalar $D_0^\ast$ SU(3) triplet, $D_0^\ast=(D_0^{\ast0},D_0^{\ast+},
D_{s0}^{\ast+})$, the corresponding Lagrangian is
\bea
\mathcal{L}_{D_0^\ast D\phi}=g_0\left(D_0^\ast u^\mu\mathcal{D}_\mu
D^\dagger+\mathcal{D}_\mu D u^\mu D_0^{\ast\dagger}\right)\ .
\eea
The coupling $g_0$ will be determined via the LO calculation of the
decay $D_0^\ast\to D^+\pi^-$.

As for the tensor $D_2^\ast$ triplet,
$D_2^\ast=(D_2^{\ast0},D_2^{\ast+},D_{s2}^{\ast+})$, the lowest-order
Lagrangian for the $D_2^\ast D\Phi$ interaction
\begin{equation}
\mathcal{L}_{D_2^\ast D\Phi}\propto
D_{2,\mu\nu}^\ast\{\mathcal{D}^\mu,u^\nu\} D^\dagger+h.c.,
\end{equation}
is of  second chiral order. Physically, the coupling of a tensor charmed
meson to a pseudoscalar charmed meson and a light pseudoscalar is in a
$D$-wave, and starts from $\order{p^2}$. Thus, the exchange of tensor charmed
mesons
will not contribute to the LECs in the $\mathcal{O}(p^2)$ and
$\mathcal{O}(p^3)$ Lagrangians, and its contribution starts from
$\mathcal{O}(p^4)$ in the chiral expansion.

The axial vector charmed mesons $D_1$ and $D_1'$ do not contribute to the LECs
in the chiral Lagrangian for the $D\phi$ interactions since there is no
$D_1^{(\prime)}D\phi$ coupling due to parity conservation. Yet, they will
contribute to those in the $D^*\phi$ Lagrangian. At leading order of the heavy
quark expansion, such contributions are equal to those of the $D_0^*$ and
$D_2^*$ to the $D\phi$ Lagrangian.

Therefore, among the excited charmed mesons, we only need to take into account
the exchange of the scalar ones for our purpose of estimating resonance
contributions to the $\mathcal{O}(p^2)$ and $\mathcal{O}(p^3)$ LECs. With the
above Lagrangians, we can then calculate the $D\phi$ scattering amplitudes by
exchanging the $D_0^*$ whose Feynman diagrams are shown in Fig.~\ref{fig:Res}
(see the first and third diagrams).
It is evident that they contribute to $D$$\phi$ scattering in both the $s$- and
$u$-channels.

For the light-flavor mesonic resonances, the low-lying vector, scalar and tensor
states will be  considered. They contribute to $D$$\phi$
scattering in $t$-channel, see the second Feynman diagram in
Fig.~\ref{fig:Res}. For the vector resonance, the involved interactions
read~\cite{Ecker:1988te,Ecker:1989yg}
\bea
\Lag_{V\phi\phi}\al=\al {i\,g_V\over{\sqrt{2}}}\langle \hat{V}_{\mu\nu}u^\mu
u^\nu\rangle\ , \qquad (\hat{V}_{\mu\nu}\equiv \partial_\mu V_\nu-\partial_\nu
V_\mu)\ , \nonumber\\
\Lag_{DDV}\al=\al i\, g_{DDV}\left\{D\,V^\mu (\partial_\mu D^\dagger)-
(\partial_\mu D)\,V^\mu D^\dagger \right\}\ ,
\label{eq:LV}
\eea
where $V_\mu$ denotes the vector meson multiplet of interest with its explicit
form given by
\bea
V_\mu=  \left( \begin{matrix}
      {\rho^0\over\sqrt{2}}+{\omega\over\sqrt{2}} & \rho^+ & K^{\ast+}\\
      \rho^- & -{\rho^0\over\sqrt{2}}+{\omega\over\sqrt{2}} &K^{*0}\\
      K^{*-}&\bar{K}^{*0}&\tilde{\phi}
   \end{matrix}\right)_\mu\ .
\eea
Here, the ideal mixing scheme between $\omega_1$ and $\omega_8$, i.e.,
$\omega_1=\sqrt{{2}/{3}}\omega+\sqrt{{1}/{3}}\tilde{\phi}$ and
$\omega_8=\sqrt{{1}/{3}}\omega-\sqrt{{2}/{3}}\tilde{\phi}$, is employed to
construct the physical $\omega$ and $\tilde{\phi}$. Note that we have denoted the
physical $\phi(1020)$ by the symbol $\tilde{\phi}$ in order to avoid
possible confusion with the notation for the matrix of Goldstone bosons.

The Lagrangians concerning the scalar resonance
exchange take the following form~\cite{Ecker:1988te,Ecker:1989yg}
\bea\label{eq:LagScalar}
\Lag_{S\phi\phi}\al=\al c_d\langle S\,u_\mu u^\mu\rangle+c_m\langle S\chi_+
\rangle+ \tilde{c}_d\,S_1\langle u_\mu u^\mu\rangle+\tilde{c}_m\,
S_1\langle\chi_+ \rangle\ , \nonumber\\
\Lag_{DDS}\al=\al g_{DDS}D\,S\,D^\dagger+ \tilde{g}_{DDS}DD^\dagger S_1\ ,
\eea
with $S_1$ and $S$ denoting the scalar singlet and octet, respectively.
In Eqs.~\eqref{eq:LV} and \eqref{eq:LagScalar}, the Lagrangians for the coupling
of the light-flavor resonances to the Goldstone bosons are taken from
Refs.~\cite{Ecker:1988te,Ecker:1989yg}.

Finally, we also consider the light-flavor tensor resonances with quantum
numbers $2^{++}$.
We collect the corresponding Lagrangians and the contribution to LECs in
Appendix~\ref{app:h5}. The involved coupling between the $D$ mesons and the
tensor resonance, $g_{DDT}$, is estimated via the method of QCD sum rules in
Appendix~\ref{app:sumrule}.

\subsection{Resonance contributions to the LECs\label{sec:contribution}}

\begin{figure}[t]
   \centering
   \includegraphics[scale=0.8]{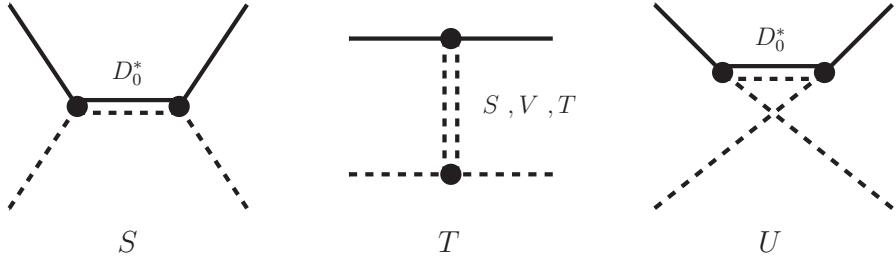}
   \caption{Diagrams for the resonance-exchange contribution to  $D$$\phi$
   scattering.
   }
   \label{fig:Res}
\end{figure}

The resonance-exchange amplitudes, corresponding to the Feynman diagrams in
Fig.~\ref{fig:Res}, are calculated and their explicit expressions are given in
Appendix~\ref{app:LOBorn} for completeness.\footnote{In Appendix~\ref{app:h5},
we will employ the technique used in Ref.~\cite{Ecker:1988te,Ecker:1989yg} to
calculate the contribution of light tensor resonances, which is different from
the one used here. } In order to calculate the tree-level
resonance-exchange contribution to the LECs, these Born-term amplitudes are
expanded in terms of $s-M_0^2$, $M_\phi^2$ and $t$ and then compared with the
contact term contributions given in Ref.~\cite{Yao:2015qia}. Consequently,
the $D_0^\ast$-exchange  contributions to the LECs are
\bea\label{eq:D0starEx}
h_5^{D_0^\ast}&=&-\frac{g_0^2}{2\Delta_0^2}\ ,\nonumber\\
g_1^{D_0^\ast}&=&g_2^{D_0^\ast}=g_3^{D_0^\ast}\,\Delta_0^2=-\frac{g_0^2}{8\Delta_0^2}\ ,
\eea
with the  difference  of the squared masses, $\Delta_0^2\equiv
\overset{\circ\;\;\;}{M_0^{\ast2}}-\overset{\circ\;}{M_0^2}$, where
$\overset{\circ\;\;\;}{M_0^{\ast}}$ is the chiral limit mass of  the $D_0^\ast$.

The light vector mesons contribute to $g_1$ and $g_2$ as follows
\bea\label{eq:vectorEx}
g_1^{V}=g_2^{V}=-\frac{g_{DDV}\,g_V}{2\sqrt{2}\,M_V^2}\ .
\eea
From the above equations, we note that light vector-meson exchange does not
 contribute to any LEC in the ${\cal O}(p^2)$ Lagrangian, which is
different from the pion-nucleon case in Ref.~\cite{Bernard:1996gq}. This is due
to the fact that the Lorentz index of the vector is contracted with the Gamma
matrices in the nucleon case, but in our case with those of the derivatives of the
$D$ mesons. For a $t$-channel exchange, this partial derivative
contributes as ${\cal O}(p^1)$. Together with the ${\cal O}(p^2)$ $V\phi\phi$
vertex  the $t$-channel vector-meson exchange thus starts to contribute at
${\cal O}(p^3)$.

The light-flavor scalar mesons contribute as
\bea\label{LECsS1S8}
h_0^{S}&=&-\frac{\tilde{g}_{DDS}\,\tilde{c}_m}{M_{S1}^2}+
\frac{{g}_{DDS}\,{c}_m}{3M_{S8}^2}\ ,\quad
h_1^{S}=-\frac{g_{DDS}\,c_m}{M_{S8}^2}\ ,\nonumber\\
h_2^{S}&=&\frac{\tilde{g}_{DDS}\,\tilde{c}_d}{M_{S1}^2}-
\frac{g_{DDS}c_d}{3M_{S8}^2}\ ,\qquad h_3^{S}=-\frac{g_{DDS}\,c_d}{M_{S8}^2}\ .
\eea
Here, $M_{S1}$ and $M_{S8}$ denote the masses of singlet and octet scalars,
respectively. Without entering the discussion about which values should be used
for the light scalar multiplets, we make use of large $N_c$  and set
$M_S=M_{S_1}=M_{S_8}$, as done in Ref.~\cite{Bernard:1996gq}. Furthermore, the
singlet couplings can be expressed in terms of the octet ones through the
relations: $\tilde{c}_{m,d}=c_{m,d}/\sqrt{3}$ and
$\tilde{g}_{DDS}=g_{DDS}/\sqrt{3}$. By imposing these large-$N_c$ relations, the
above expressions in Eqs.~(\ref{LECsS1S8}) are reduced to
\bea\label{eq:scalarEx}
h_0^S=0\ ,\quad h_1^{S}=-\frac{g_{DDS}\,c_m}{M_{S}^2}\ ,\quad h_2^S=0\ ,\quad
h_3^S=-\frac{g_{DDS}\,c_d}{M_{S}^2}\ .
\eea
One sees that the LECs $h_0$ and $h_2$ receive no contribution from the light
scalar mesons in the large-$N_c$ limit. In fact, these two LECs, together with
$h_4$, are of one order higher in the $1/N_c$ expansion in comparison with
$h_i(i=1,3,5)$~\cite{Lutz:2007sk,Guo:2008gp}.

As shown in Appendix~\ref{app:h5}, the exchange of light tensor mesons with
$J^{PC}=2^{++}$ contributes only to $h_5$, which is of the form
\bea\label{eq:tensorEx}
h_5^{T}=\frac{g_{DDT}g_T}{M_T^2}\ ,
\eea
with $g_{DDT}$ and $g_T$  the coupling constants for $D$-$D$-tensor and
$\pi$-$\pi$-tensor vertices, respectively, see Eqs.~\eqref{lddt} and
(\ref{eq:LagpipiT}). This contribute was calculated employing the technique used in
Refs.~\cite{Ecker:1988te,Ecker:1989yg}, namely matching the effective actions,
 which is equivalent to the approach we
have used above that is based on the matching using the explicit perturbative amplitudes in both theories.

\subsection{Numerical results\label{sec:contributionnum}}

To obtain numerical estimates for the LECs, we need to know the resonance
couplings. However, not all of the involved couplings are really known. Thus,
for the measurable ones ($\tilde g,g_0,g_V,c_d,c_m$ and $g_T$), we will
extract the values from experimental data, and for the ones in the vertices
where not all three particles can on go shell simultaneously
($g_{DDV},g_{DDS}$ and $g_{DDT}$), we will take model values for an estimate.

The numerical value of the resonance coupling $g_0$ can be obtained by
calculating the decay width $\Gamma({D_0^\ast\to D^+\pi^-})$. At LO, we have
\begin{equation}
\Gamma({D_0^\ast\to D^+\pi^-})
=\frac{1}{4\pi}\frac{|g_0^2|}{F_0^2}\frac{\left(m_{D_0^\ast}\sqrt{
M_\pi^2+|\vec{q}_\pi|^2}-M_\pi^2\right)^2}{m_{D_0^\ast}^2} ,
\end{equation}
with $\vec q_\pi$ the pion momentum in the rest frame of the initial particle.
Comparing with the empirical value taken from  the Particle Data
Group~\cite{Agashe:2014kda}, we get
\begin{equation}
|g_0|=0.68\pm0.05\, .
\end{equation}
The couplings of the light-flavor resonances to the Goldstone bosons, $g_V$ and
$c_{m,d}$, have been used in many studies of the chiral resonance
Lagrangian, and we take the updated determinations in Ref.~\cite{Guo:2009hi}
\begin{equation}
|g_V|=0.0846\pm 0.0008\, ,\qquad
|c_m|=(80\pm 21)~{\rm MeV}\, ,\qquad
|c_d|=(26\pm 7)~{\rm MeV}\,.
\end{equation}
From the decay width of the $f_2(1270)\to\pi\pi$~\cite{Agashe:2014kda}, we get
$|g_T|=28$~MeV. Note that, as discussed in Ref.~\cite{Bernard:1996gq}, if the
$\pi$$N$ LEC $c_1$ is completely saturated by scalar exchange, a positive $c_m$
is demanded.  Together with the constraint $4 c_m c_d=F_0^2$, see e.g.
Ref.~\cite{Jamin:2001zq}, we will set $c_{m,d}>0$ in the following.

For the troublesome couplings, we take the following values:
\bea
g_{DDV}=1.46\,,\qquad
g_{DDS}=5058~{\rm MeV}\, ,\qquad
g_{DDT}=3.9\times 10^{-3}~{\rm MeV}^{-1}\, ,
\label{eq:trouble}
\eea
The value of $g_{DDV}$ is taken from the analysis of the $DDV$ vertex using
light-cone QCD sum rules~{\cite{Wang:2007mc}}. For the $g_{DDS}$, we have
utilized the large-$N_c$ relation $g_{DDS}={\sqrt{3}}\tilde{g}_{DDS}$, and take
the value $g_{DD\sigma}$ used in Ref.~\cite{Ding:2008gr}, which was extracted
from the parity doubling model of Ref.~\cite{Bardeen:2003kt}, for  $\tilde
g_{DDS}$.
There is no available modeling of $g_{DDT}$, and we thus estimate it using QCD
sum rules in Appendix~\ref{app:sumrule}. The problem  is that it is hard to
quantify the uncertainty of these parameters. Yet, there is evidence that
these model values are of the right order: the dimensionless values for
$g_{DDV}$, $g_{DDS}/\Lambda_\text{had}\sim5$ and $g_{DDT}
\Lambda_\text{had}\sim 4$, where $\Lambda_\text{had}=\order{1~\text{GeV}}$ is
a typical hadronic scale, have more or less natural sizes of $\order{1}$.

For the masses involved in our numerical estimate, we take
\bea
\overset{\circ}{M}\al\cong\al\overline{M}_D=\frac{1}{2}(M_D^{\rm
phy.}+M_{D_s}^{\rm phy.})=1918~{\rm MeV}\, ,\nonumber\\
\overset{\circ\;\;}{M_0^{\ast}}\al\cong\al\overline{M}_{D_0^\ast}=
\frac{1}{2}(M_{D_0^\ast}^{\rm phy.}+ M_{D_{s0}^\ast}^{\rm phy.})=2318~{\rm
MeV}\, ,\nonumber\\
M_V\al=\al764~{\rm MeV}\, ,\quad
M_S = 980~{\rm MeV}\, , \quad
M_T =  1270~{\rm MeV}\, ,
\eea
where the chiral limit masses are identified with  the corresponding averaged
physical masses, which is acceptable given  the accuracy we are aiming at.
To be consistent with  using the values of $g_V$ and $c_{m,d}$
given above, the values for $M_V$ and $M_S$ are also taken from
Ref.~\cite{Guo:2009hi}. The mass for the tensor multiplet is chosen to be the
mass of $f_{2}(1270)$ following Ref.~\cite{Ecker:2007us}.

With the resonance couplings and masses specified above, we are now in the
position to estimate the resonance contributions to the LECs based on the
analytical expressions, Eqs.~(\ref{eq:D0starEx}-\ref{eq:tensorEx}). The
numerical results are shown in Table~\ref{tab:LECsRes} and the sum of various
contributions is given in the last column. Because of the poor knowledge on the
values of the off-shell couplings $g_{DDV,DDS,DDT}$, no reasonable error
estimate can be made here.
Furthermore, the signs of $g_{V,T}$ are not fixed, hence contributions from the
$t$-channel exchanges of the light-flavor vector and tensor mesons might be
either positive or negative as listed in Table~\ref{tab:LECsRes}, and they also
take two possible values in the last column of the table due to interference
with the contribution from the scalar charmed mesons.

\begin{table}[t]
\caption{Estimates of the resonance contributions to the LECs. Here $h_{0,2,4}$,
which vanish in the large-$N_c$ limit, are not shown. The columns starting with $D_0^*$, $V$, $S$ and $T$ 
list the contributions from the exchange of the scalar charmed mesons, light-flavor vector, scalar and
tensor mesons, respectively. The last column sums over all these contributions. }
\label{tab:LECsRes}
\begin{center}
\newcolumntype{C}{>{\centering\arraybackslash}X}
\newcolumntype{R}{>{\raggedleft\arraybackslash}X}
\renewcommand{\arraystretch}{1.5}
\begin{tabularx}{0.8\textwidth}{lCCCCC}
\hline\hline
LEC		& $D_0^\ast$&$V$	&$S$  &$T$& ${\text{Total}}$\\
\hline
$h_1$  	&0	&0	&$0.4$	&0&$0.4$\\
$h_3$ 	&0	&0	&$0.1$	&0&$2.3$\\
$h_5$ [GeV$^{-2}$]	&$-0.1\phantom{0}$	&0	&0	&$\pm0.1$ &$[-0.5,-0.3]$\\
$g_1$ [GeV$^{-2}$]		&$-0.03$&$\mp0.07$	&0	&0&$[-0.04,0.1]$\\
$g_2$ [GeV$^{-2}$]		&$-0.03$&$\mp0.07$	&0	&0&$[-0.04,0.1]$\\
$g_3$ [GeV$^{-4}$]		&$\phantom{-}0.02$&0	&0	&0&$0.02$\\
\hline\hline
\end{tabularx}
\end{center}
\end{table}

\subsection{Comparison with results from unitarized ChPT (UChPT)}

We compare the estimates of the LECs with those from fits to the lattice data on
of scattering lengths of some selected channels at the NLO and NNLO in the
framework of UChPT~\cite{Oller:2000fj} (and references therein) in
Table~\ref{tab:LECsP2} and Table~\ref{tab:LECsP3},
respectively, where the definitions of the combinations of the LECs are given
in Eq.~(\ref{eq:LECsCom}).\footnote{The notations of the LECs adopted in
Ref.~\cite{Altenbuchinger:2013vwa} are connected to ours by $h_0=2c_0\ ,\,h_1=-2c_1\
,$$$\,h_{24}=2\left[c_{24}+2c_4(1-\frac{\bar{M}_D^2}{m_P^2})\right]\
,\,h_{35}=-2\left[c_{35}+2c_5(1-\frac{\bar{M}_D^2}{m_P^2})\right]\
,\,h_4^\prime=-4c_4\frac{\bar{M}_D^2}{m_P^2}\ ,\,
h_5^\prime=2c_5\frac{\bar{M}_D^2}{m_P^2}\ ,$$ with $m_P=1.9721$~GeV specified in
Ref.~\cite{Altenbuchinger:2013vwa} and $\bar{M}_D=(M_D^{\rm phy}+M_{D_s}^{\rm
phy})/2$.} The following observations can be made:

\begin{enumerate}

\item[(i)]{} Provided that a positive value of $c_m$ is chosen, $h_1$ is saturated by the light scalar exchange, which is similar
to the LEC $c_1$ in the $\pi N$ case~\cite{Bernard:1996gq}. The value of $h_1$
is fixed through the mass difference between strange and nonstrange charmed
mesons, which is then adopted in these fits. One sees that the estimate here is
in a good  agreement with the empirical value. The agreement in turn might
indicate that the model estimate for $g_{DDS}$ is reasonable.

\item[(ii)]{} Because we only have the absolute values for $g_V$ and $g_T$,
one sees that the estimates from exchanging
resonances are roughly consistent with those determined from the various NLO
UChPT fits, while there are tensions when comparing with those from the NNLO
UChPT fits in Ref.~\cite{Yao:2015qia}. While there are quite a few fit
parameters at NNLO, not many lattice data exist. On the one hand, more lattice
calculations on observables for the scattering processes between heavy mesons
and light mesons would be welcome to better pin down the LECs at NNLO. On the
other hand, as pointed out in Ref.~\cite{Ecker:1988te}, the values of the LECs,
which are scale-dependent in general, are dominated by the resonances only when
the renormalization scale $\mu$ is not too far away from the resonance region.
The NLO fit results are obtained with $\mu=1$~GeV, i.e. the scale appearing in the subtraction constant $a(\mu)$, 
which is around the masses of the light vector and scalar resonances. 
However, the NNLO fit results in Ref.~\cite{Yao:2015qia} are obtained by 
setting $\mu=\bar{M}_D=1.92$~GeV for convenience. Note that the scale-dependence of the NNLO LECs stems both from the renormalization of the one-loop amplitude using EOMS scheme and  the unitarization procedure accompanied by the subtraction constant $a(\mu)$.

\end{enumerate}

\begin{table}[t]
\caption{Comparison of the values of the LECs from the estimate
using resonances with those from fits to lattice data in various formulations of
unitarized ChPT at NLO. The LECs in this table are dimensionless.
}\label{tab:LECsP2}
\begin{center}
\newcolumntype{C}{>{\centering\arraybackslash}X}
\newcolumntype{R}{>{\raggedleft\arraybackslash}X}
\renewcommand{\arraystretch}{1.5}
\begin{tabularx}{\textwidth}{CCCCCC}
\hline\hline
LEC	&Table V~\cite{Liu:2012zya}	& Table VIII~\cite{Liu:2012zya}& HQS~\cite{Altenbuchinger:2013vwa}	& $\chi$-SU(3)~\cite{Altenbuchinger:2013vwa}  & Resonance\\
\hline
$h_0$	&0.01		&0.01		&0.03		&0.03	&0\\
$h_1$	&0.42		&0.42		&0.43		&0.43	&$0.4$\\
$h_{24}$	&$-0.10_{-0.06}^{+0.05}$&$0.10_{-0.06}^{+0.05}$&$-0.12\pm0.05$	&$-0.14\pm0.04$&0\\
$h_{35}$
&$\phantom{-}0.25_{-0.13}^{+0.13}$&$0.26_{-0.10}^{+0.09}$&
$\phantom{-}0.23\pm0.09$ & $\phantom{-}0.12\pm0.08$&$[-1.4,0.1]$\\
$h_4^\prime$	&$-0.32_{-0.34}^{+0.35}$&$-0.30_{-0.28}^{+0.31}$&$-0.20\pm0.31$	&$-0.83\pm0.30$&0\\
$h_5^\prime$	&$-1.88_{-0.61}^{+0.63}$&$-1.94_{-0.38}^{+0.46}$&$-1.82\pm0.57$	&$-1.00\pm0.40$&$[-0.7,0]$\\
\hline\hline
\end{tabularx}
\end{center}
\end{table}

\begin{table}[t]
\caption{
Comparison of the values of the LECs from the estimate using resonances
with those from various fits to lattice data in unitarized ChPT at
NNLO.}\label{tab:LECsP3}
\begin{center}
\newcolumntype{C}{>{\centering\arraybackslash}X}
\newcolumntype{R}{>{\raggedleft\arraybackslash}X}
\renewcommand{\arraystretch}{1.5}
\begin{tabularx}{\textwidth}{lCCCCC}
\hline\hline
LEC	&U$\chi$PT-6a~\cite{Yao:2015qia}	& U$\chi$PT-6b~\cite{Yao:2015qia}& U$\chi$PT-6$a^\prime$~\cite{Yao:2015qia}	& U$\chi$PT-6$b^\prime$~\cite{Yao:2015qia} & Resonance\\
\hline
$h_0$	&0.02		&0.02		&0.02		&0.02	&0\\
$h_1$	&0.43		&0.43		&0.43		&0.43	&$0.4$\\
$h_{24}$  & $\phantom{-}0.79_{-0.09}^{+0.10}$	&
$\phantom{-}0.76_{-0.09}^{+0.10}$	& $\phantom{-}0.83_{-0.10}^{+0.11}$	&
$\phantom{-}0.80_{-0.10}^{+0.10}$&0\\
$h_{35}$	&$ \phantom{-}0.73_{-0.38}^{+0.50}$	& $\phantom{-}0.81_{-0.62}^{+0.95}$
& $\phantom{-}0.43_{-0.23}^{+0.23}$	&
$\phantom{-}0.40_{-0.29}^{+0.33}$&$[-1.4,0.1]$\\
$h_4^\prime$ & $-1.49_{-0.57}^{+0.55}$& $-1.56_{-0.65}^{+0.61}$&$ -1.33_{-0.60}^{+0.60}$&$ -1.72_{-0.63}^{+0.64}$&0\\
$h_5^\prime$ & \!\!\!$-11.47_{-2.79}^{+2.24}$& \!\!\!$-15.38_{-7.20}^{+4.81}$	&
$-4.25_{-0.66}^{+0.65}$ &$ -2.60_{-0.87}^{+0.84}$&$[-0.7,0]$\\
$g_{1}^\prime\,[\text{GeV}^{-1}]$& $-1.66_{-1.59}^{+0.31}$ 	&
$-2.44_{-0.64}^{+0.57}$ & $-1.10_{-0.23}^{+0.18}$	& $-1.90_{-0.35}^{+0.58}$&$[-0.2,-0.1]$\\
$g_{23}\,[\text{GeV}^{-1}]$ 	& $-1.24_{-1.51}^{+0.28}$ 	&$
-2.00_{-0.51}^{+0.52}$  & $-0.70_{-0.24}^{+0.19}$  & $-1.48_{-0.37}^{+0.61}$&$[-0.5,-0.2]$\\
$g_3^\prime\,[\text{GeV}^{-1}]$ &$ \phantom{-}2.12_{-0.45}^{+0.55}$ 	&
$\phantom{-}2.85_{-0.96}^{+1.41}$	& $\phantom{-}0.98_{-0.14}^{+0.15}$  &
$\phantom{-}0.58_{-0.19}^{+0.20}$&$0.14$\\
\hline\hline
\end{tabularx}
\end{center}
\end{table}

\section{Positivity constraints on the $D$$\phi$ interactions\label{SecPos}}

In this section,  positivity constraints on the $D$$\phi$ interactions will be
derived by using basic axiomatic principles of $S$-Matrix theory, such as
unitarity, analyticity and crossing symmetry. Such constraints are important in
the sense that model-independent information for the $D$$\phi$ interactions is
provided. When employed in ChPT, they are translated into a much more practical
form, i.e. positivity bounds on the LECs. In general, these involved LECs are
unknown and not fixed by chiral symmetry. Furthermore, the number of the LECs
increases when going to higher orders. Therefore,  such bounds are of great use,
especially for those which can not be measured directly in experiments such as
the $D$$\phi$ interactions under consideration.
In what follows, details on the derivation of these constraints as well as practical
applications of the bounds on the $D$$\phi$ interactions will be presented.

\subsection{Positivity constraints implied by dispersion relations}

For elastic $D$$\phi$ scattering, the Mandelstam triangle is the region bounded
by $s=(M_D+M_\phi)^2$, $u=(M_D+M_\phi)^2$ and $t=4M_\phi^2$ in the Mandelstam
plane as displayed in Fig.~\ref{fig:Mandelstam}. Inside the Mandelstam triangle, the
scattering amplitude is analytic and real, see e.g.
Ref.~\cite{Buettiker:1999ap} for an early application in the context baryon ChPT. Following
Refs.~\cite{Manohar:2008tc,Sanz-Cillero:2013ipa}, we restrict ourselves to
the upper part of the Mandelstam triangle with $t\geq0$. Using unitarity,
analyticity and crossing symmetry, an $n$-time subtracted fixed-$t$ dispersion
relation for the elastic $D$$\phi$ scattering amplitude with definite $(S,I)$
can be written as
\bea
\frac{\text{d}^n}{\text{d}s^n}\mathcal{M}^{(S,I)}_{D\phi\to D\phi}(s,t) \al=\al
\frac{n!}{\pi}\int_{(M_D+M_\phi)^2}^{+\infty}\text{d}x^\prime
\Bigg[ \delta^{I I'}
\frac{\text{Im}\,\mathcal{M}^{(S,I')}_{D\phi\to D\phi}
(x^\prime+i\epsilon,t)}{(x^\prime-s)^{n+1}} \nonumber\\ \al\al +
(-1)^nC_{us}^{II^\prime}\,\frac{\text{Im}\,
\mathcal{M}^{(S,I^\prime)}_{D\bar{\phi}\to D\bar{\phi}}(x^\prime+i\epsilon,t)
}{(x^\prime-u)^{n+1}} \Bigg]\ ,
\label{eqDR}
\eea
where $\bar{\phi}$ denotes the antiparticle of $\phi$,  and
$C_{us}^{II^\prime}$ represents the $u$-$s$ crossing matrix which is defined
as
\be
 \mathcal{A}^I(u,t,s) = C_{us}^{II'} \mathcal{A}^{I'}(s,t,u),
\ee
where we have written explicitly all of the three Mandelstam variables so as to
make the $u$-$s$ crossing explicit, and $C_{su}$ is defined by exchanging
the $s$- and $u$-channel amplitudes in the above equation.
The matrices satisfy $C_{us}^{II'} C_{su}^{I'J}=\delta^{IJ}$.  We want to
mention that the imaginary part of $\mathcal{M}$ is positive definite above
threshold.\footnote{Here, we follow the convention
$S=\bold{1}+i\,(2\pi)^4\delta^{(4)}(\sum_i p_i-\sum_f p_f){\cal M}\ ,$
 to define the scattering amplitude ${\cal M}(s,t)$.}
Besides, we have assumed that all
the processes involved in the dispersion relation are single-channel
interactions such that the integration starts at the corresponding thresholds.
The case with multi-channel interactions will be discussed later. Both imaginary
parts in the brackets in Eq.~(\ref{eqDR}) are positive definite when $x^\prime$
is above threshold, i.e.
$x^\prime>(M_D+M_\phi)^2$.\footnote{ Since for each partial wave $\ell$, ${\rm
Im\ }
\mathcal{M}_\ell(s)=\frac{2|\vec{k}|}{\sqrt{s}}|\mathcal{M}_{\ell}(s)|^2\geq0$
above threshold and the Legendre polynomials $P_\ell(\cos\theta)\geq0$ for
$t\geq0$ (or equivalently $\cos\theta\geq 1$), one has ${\rm
Im}\,\mathcal{M}^{(S,I)}(s,t)=\sum_{\ell=0}^\infty(2\ell+1)\,P_\ell(t)\,{\rm
Im}\mathcal{M}_\ell^{(S,I)}(s)\geq0$.} In addition, the $s$-channel coefficient
$\delta^{II^\prime}$ is always non-negative. However, the $u$-channel one
$(-1)^nC_{us}^{II^\prime}$ is  sometimes not.
The aim is therefore  to construct certain combinations of the $D$$\phi$
amplitudes with different isospins such that
\bea
\frac{{\rm d}^n}{{\rm d}s^n} \left[\alpha^I\mathcal{M}^{(S,I)}_{D\phi\to
D\phi}(s,t)\right]\geq0\ .\label{eqPosCon}
\eea
where summation over  $I$ is assumed.
In combination with Eq.~(\ref{eqDR}), a sufficient condition for the above
positivity condition to hold is given by
\bea
 \alpha^I \delta^{I I'} \geq0 \ ,\quad
      \alpha^IC_{us}^{II^\prime}\geq0 \  (\text{for even $n$})\
      .\label{eqCombi}
\eea

For the multi-channel case, all the cuts from the coupled channels need to be
taken into account, and the integration should start from the lowest
threshold.
Taking the process $DK\to DK$ as an example, the integration in the $(S,I)=(1,1)$
channel will start at the $D_s\pi$ threshold rather than its physical $DK$ threshold.
Since the imaginary part $\text{Im}\,T^{(1,1)}_{DK\to
DK}(x^\prime+i\epsilon,t)$ could be negative in the region
$x^\prime\in\left[(M_{D_s}+M_\pi)^2,(M_D+M_K)^2\right]$, the positivity
condition in Eq.~(\ref{eqPosCon}) is not applicable any more.
However, as discussed in Ref.~\cite{Mateu:2008gv}, in the multi-channel case,
the positivity conditions hold for processes of the type $a+b\to a+b$ such that
$m_a+m_b$ is the lightest threshold for both the $s$- and $u$-channels. This
statement is obtained from the condition that the dispersion relation in
Eq.~(\ref{eqDR}) is true and that $t\geq0$ which ensures the positivity of the
Legendre polynomials for all partial waves.
For details we refer to Section IV in Ref.~\cite{Mateu:2008gv}. With this
statement, amongst all the $D\phi$ scattering channels, only $D\pi\to D\pi$ and
$D_s\pi\to D_s\pi$ survive.

In order to derive positivity constraints on the $D\pi\to D\pi$ and $D_s\pi\to
D_s\pi$ scattering amplitudes, we need to know the explicit forms of the
$u$-$s$ crossing matrices for these two processes, which are 
\bea
\mathcal{C}_{us}=\begin{pmatrix}
    -\frac{1}{3}  & \frac{4}{3}   \\
    \frac{2}{3}  & \frac{1}{3}
\end{pmatrix} \ ,\quad \text{for }D\pi\to D\pi\ , \quad \text{and} \quad
\mathcal{C}_{us}=1\ ,\quad\text{for }D_s\pi\to D_s\pi \ .
\eea
For the $D\pi$ case, the matrix is arranged such that the first channel refers
to $I=1/2$ and the second $I=3/2$.
From now on, we will focus on the $n=2$ case, which is the minimal number of
subtractions in the dispersion integral required by the Froissart
bound~\cite{Froissart:1961ux}.

For {$D\pi\to D\pi$}, the upper-part of the Mandelstam triangle is
$\mathcal{R}_{D\pi}=\{(s,t)|s\leq(M_D+M_\pi)^2,s+t\geq(M_D-M_\pi)^2,0\leq t\leq 4M_\pi^2\}$.
When $(s,t)\in\mathcal{R}_{D\pi}$, a sufficient condition for ${\frac{{\rm
 d}^2}{{\rm d}s^2}\left\{\alpha^I\mathcal{M}^I_{D\pi\to
D\pi}(s,t)\right\}\geq0}$ is given by $2\alpha^{3/2}\geq\alpha^{1/2}\geq0$. We
choose the following three combinations of $\alpha^{1/2}$ and $\alpha^{3/2}$ to
get bounds on three physical scattering amplitudes:
\bea
  \left\{ \begin{array}{ccc}
      \alpha^{1/2}=0\ , \alpha^{3/2}=1\ : && -\frac{{\rm d}^2}{{\rm
d}s^2}\mathcal{A}_{D^+\pi^+\to D^+\pi^+}(s,t)\geq0\ ,\\
      \alpha^{1/2}=\frac{2}{3}\ , \alpha^{3/2}=\frac{1}{3}\ : && -\frac{{\rm
d}^2}{{\rm d}s^2}\mathcal{A}_{D^0\pi^+\to D^0\pi^+}(s,t)\geq0\ ,\\
        \alpha^{1/2}=\frac{1}{3}\ , \alpha^{3/2}=\frac{2}{3}\ : && -\frac{{\rm
d}^2}{{\rm d}s^2}\mathcal{A}_{D^+\pi^0\to D^+\pi^0}(s,t)\geq0\ ,\\
   \end{array}\right.
   \label{eq:posDpi}
\eea
with $\mathcal{A}=-\mathcal{M}$.

For $D_s\pi\to D_s\pi$, the upper-part of the Mandelstam triangle is
$\mathcal{R}_{D_s\pi}=\{(s,t)|s\leq(M_{D_s}+M_\pi)^2,s+t\geq(M_{D_s}-M_\pi)^2,
0\leq t\leq 4M_\pi^2\}$.  When $(s,t)\in\mathcal{R}_{D_s\pi}$, a sufficient
condition for ${\frac{{\rm d}^2}{{\rm
d}s^2}\left\{\alpha^I\mathcal{M}^I_{D_s\pi\to D_s\pi}(s,t)\right\}\geq0}$ is
$\alpha^{1}\geq0$. Choosing $\alpha^{1}=1$,
one has
\bea
-\frac{{\rm d}^2}{{\rm d}s^2}{ \mathcal{A}}_{D_s^+\pi^+\to
D_s^+\pi^+}(s,t)\geq0\ .
\label{eq:posDspi}
\eea

In the above, we have written the constraints in terms of the scattering
amplitudes which are either explicitly given in Ref.~\cite{Yao:2015qia} or
easily obtainable by using crossing symmetry and isospin symmetry. Hence their
analytical expressions up to NNLO are all known and can be inserted into the
above inequalities to obtain bounds on the LECs, which will be discussed in the
next section.

\subsection{Positivity bounds on the LECs\label{LECbound}}

The representation of the $D$$\phi$ scattering amplitudes in the manifestly
Lorentz covariant framework obtained in Ref.~\cite{Yao:2015qia} is suitable to
obtain reliable bounds on the LECs, since it possesses the correct analytic
behavior inside the Mandelstam triangle. In the covariant formalism for the
$SU(3)$ case, the NLO (tree-level) $D$$\phi$ amplitudes were first given by
Ref.~\cite{Guo:2009ct} and then followed by
Refs.~\cite{Liu:2012zya,Wang:2012bu,Altenbuchinger:2013vwa}.
In Ref.~\cite{Yao:2015qia}, a complete covariant calculation up to NNLO (the
leading one-loop order) is presented using the EOMS subtraction scheme which
guarantees  proper analyticity and has the correct power counting. These
amplitudes can be employed to derive positivity bounds on the LECs with the help
of the inequalities given in Eq.~\eqref{eq:posDpi} and Eq.~\eqref{eq:posDspi}.
Note that, throughout this work, we follow the notations of
Ref.~\cite{Liu:2012zya}, and the results from other works with different
notations can be easily adopted to ours.

\subsubsection{Bounds up to $\mathcal{O}(p^2)$}

\begin{figure}[t]
\begin{center}
\epsfig{file=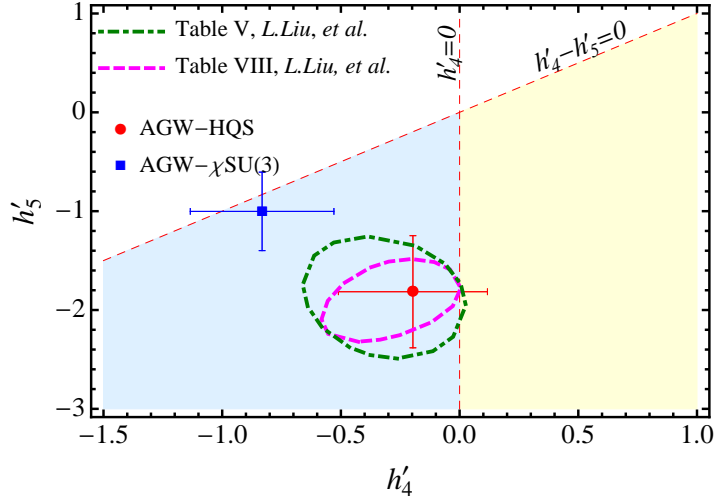,scale=1.}
\end{center}
\vspace{-0.5cm}
\caption{Comparison of the NLO positivity bounds for $h_4^\prime$ and
$h_5^\prime$ with their values obtained from fitting to the lattice data using
unitarized ChPT at NLO.
The positivity-bound region is depicted in light yellow bounded by the lines
$h_4^\prime=0$ and $h_4^\prime-h_5^\prime=0$.
The area in light blue denotes the region where the bound
$h_4^\prime-h_5^\prime\geq0$ is respected while $h_4^\prime\geq0$ is violated.
The green dot-dashed and magenta dashed ellipses represent the 1-$\sigma$
regions for $h_4^\prime$ and $h_5^\prime$ from the 5- and 4-parameter fits in
Ref.~\cite{Liu:2012zya}, respectively. The red dot and blue square with error
bars, denoted by AGW-HQS and AGW-$\chi$SU(3), respectively, are taken from
Ref.~\cite{Altenbuchinger:2013vwa}.
}
\label{fig:boundP2}
\end{figure}

Inserting the  amplitudes up to NLO into Eqs.~\eqref{eq:posDpi} and
\eqref{eq:posDspi}, the constraints on the scattering amplitudes turn into
bounds on the LECs $h_4$ and $h_5$.  Each inequality leads to one bound on the
LECs. The intersection of all of the obtained bounds  has a simple form \bea
 \left\{ \begin{array}{rcl}
      h_4-h_5 \al \geq \al 0 \\
      h_4 \al\geq \al 0\\
   \end{array}\right. , \qquad\text{or equivalently}\qquad\left\{
   \begin{array}{rcl}
      h_4^\prime-h_5^\prime \al \geq \al 0 \\
      h_4^\prime \al \geq\al 0\\
   \end{array}\right.  .\label{PosboundsNLO}
\eea
Here, the parameters $h_4$ and $h_5$ are in units of $\text{GeV}^{-2}$, while
$h_4^\prime$ and $h_5^\prime$, defined in Eq.~(\ref{eq:LECsCom}), are
dimensionless. The region restricted by the bounds on $h_4^\prime$
and $h_5^\prime$ in Eq.~(\ref{PosboundsNLO}) is depicted as the light yellow
area in Fig.~\ref{fig:boundP2}. Two different sets of  fitting values from
Refs.~\cite{Liu:2012zya,Altenbuchinger:2013vwa} which resum the NLO scattering
amplitudes in different ways are shown for comparison:\footnote{The results in
Ref.~\cite{Wang:2012bu} are not taken into consideration, since the preliminary
lattice data~\cite{Liu:2008rza}, which are different from the final ones in
Ref.~\cite{Liu:2012zya}, are used to perform fits there.}

\begin{enumerate}
\item[(i)]
The first set is taken from Ref.~\cite{Liu:2012zya}. There are two different
fits:
one with 5 parameters which are four LECs and one subtraction constant used to
regularize the loop integral (cf. Table~V therein), and the other with 4
parameters with the subtraction constant fixed from reproducing the
$D_{s0}^*(2317)$ mass in the $(S,I)=(1,0)$ channel (cf. Table~VIII therein).
The 1-$\sigma$ regions, with the parameter correlations in the fits taken into
account, from these two fits for the values for $h_4^\prime$ and $h_5^\prime$
are shown by the regions surrounded by the green dot-dashed line (for the
5-parameters fit) and by magenta dashed line (for the 4-parameter fit).

\item[(ii)]
The second set is taken from Ref.~\cite{Altenbuchinger:2013vwa}.
In that work, a special renormalization scheme is proposed to deal with the
so-called power counting breaking terms appearing in the loop functions.
In Fig.~\ref{fig:boundP2}, the blue square and red dot represent the fit values
taken from $\chi$-SU(3) fit and HQS fit, which correspond to different
regularizations of the scalar two-point scalar loop integral, in
Ref.~\cite{Altenbuchinger:2013vwa}, respectively.

\end{enumerate}

As can be seen from Fig.~\ref{fig:boundP2}, the fit values from Ref.~\cite{Liu:2012zya}
are only marginally consistent with the region allowed by the bounds. The LEC
values from the HQS fit~in Ref.~\cite{Altenbuchinger:2013vwa} has a small
overlap with the positivity bound, while the ones from the $\chi$-SU(3) fit are
completely outside the  region derived from positivity.

\subsubsection{Bounds up to $\mathcal{O}(p^3)$}
\label{sec:op3}

 \begin{figure}[th]
\begin{center}
\epsfig{file=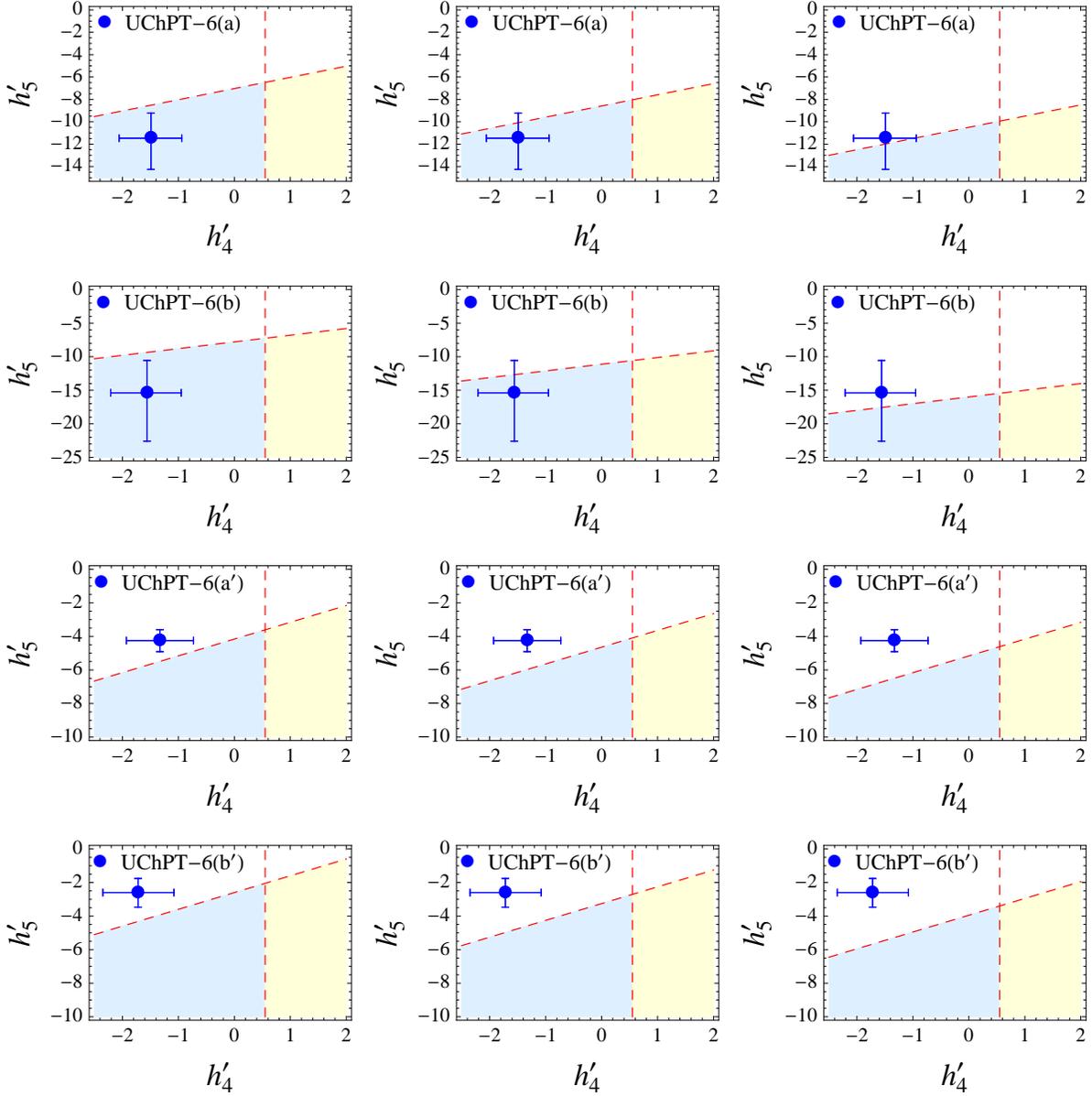,scale=0.9}
\end{center}
\vspace{-0.5cm}
\caption{Comparison of the positivity bounds for $h_4^\prime$ and
$h_5^\prime$ with their 6-channel NNLO fit values. The graphs in the first,
second and third column correspond to the case that $g_3^\prime$ is fixed at
its lowest, central and largest value, respectively. The blue dots with error
bars represent the fitting values of $h_4^\prime$ and $h_5^\prime$ from
different fits: UChPT-6(a), UChPT-6(b), UChPT-6($a^\prime$) and
UChPT-6($b^\prime$), see Ref.~\cite{Yao:2015qia}. The NNLO positivity-bound
region is  in light yellow bounded by the lines $h_4^\prime=0.55$ and
$h_4^\prime-h_5^\prime= g(g_3^\prime)$. The area in light blue denotes the
region where the bound $h_4^\prime-h_5^\prime\geq g(g_3^\prime)$ is respected
while $h_4^\prime\geq0.55$ is violated.}
\label{fig:boundP3}
\end{figure}

Inserting the $D$$\phi$ amplitudes up to NNLO into the positivity constraints
in Eqs.~\eqref{eq:posDpi} and \eqref{eq:posDspi},
one gets bounds on the LECs at the NNLO level, which are
\bea
\left\{
\begin{array}{rcll}
h_4-h_5-24M_D\,\nu_D\,g_3 \al \geq \al f^{(2)}_{D^+\pi^+\to D^+\pi^+}(s,t), \al
\quad (s,t)\in\mathcal{R}_{D\pi}\ ,\\
h_4-h_5+24M_D\,\nu_D\,g_3 \al \geq \al f^{(2)}_{D^0\pi^+\to D^0\pi^+}(s,t), \al
\quad (s,t)\in\mathcal{R}_{D\pi}\ ,\\
h_4-h_5 \al \geq \al f^{(2)}_{D^+\pi^0\to D^+\pi^0}(s,t), \al \quad
(s,t)\in\mathcal{R}_{D\pi}\ ,\\
h_4 \al \geq \al f^{(2)}_{D_s^+\pi^+\to D_s^+\pi^+}(s,t), \al \quad
(s,t)\in\mathcal{R}_{D_s\pi}\ ,
\end{array}
\right.
\eea
where
\[
\nu_D\equiv \frac{s-u}{4M_D}, \qquad \text{and}~~f^{(2)}_{\rm
process}(s,t)\equiv \frac{F_\pi^2}{2}\frac{\rd^2}{\rd s^2} \mathcal{A}_{\rm
process}^{\rm loop}(s,t).
\]
Each bound would become more
stringent if one always sets $f^{(2)}_{\rm process}(s,t)$ at its maximum inside
$\mathcal{R}_{D\pi}$ (or $\mathcal{R}_{D_s\pi}$). Numerically, we find
\begin{eqnarray}
  \max\{f^{(2)}_{D^+\pi^+\to D^+\pi^+}(s,t)\} \al=\al \max\{f^{(2)}_{D^0\pi^+\to
D^0\pi^+}(s,t)\}=0.34 , \nonumber\\
  \max\{f^{(2)}_{D^+\pi^0\to D^+\pi^0}(s,t)\} \al=\al 0.28
\end{eqnarray}
in the region $(s,t)\in\mathcal{R}_{D\pi}$, and
\begin{equation}
 \max\{f^{(2)}_{D_s^+\pi^+\to D_s^+\pi^+}(s,t)\}=0.15
\end{equation}
in the region $(s,t)\in\mathcal{R}_{D_s\pi}$. By further using the condition
$|\nu_D|\leq \nu_D^\text{th}(t)=M_\pi+t/{4M_D}\leq M_\pi+
{M_\pi^2}/{M_D}$, one finally obtains the following bounds
\bea
\left\{
\begin{array}{r}
h_4^\prime-h_5^\prime-24|g_3^\prime|(M_D+M_\pi)M_\pi/\bar{M}_D\geq 1.25\ ,\\
h_4^\prime\geq0.55\ ,
\end{array}
\right.\label{Boundp3}
\eea
which are expressed in terms of $h_4^\prime$, $h_5^\prime$ and $g_3^\prime$.
Comparing with the $\mathcal{O}(p^2)$ bounds given
in~Eq.~(\ref{PosboundsNLO}), the $\mathcal{O}(p^3)$ bounds are much more
stringent.

In order to compare the values of $h_4^\prime$ and $h_5^\prime$ from the fits
to the lattice data using unitarized ChPT at NNLO with these bounds, we choose
to fix $g_3^\prime$ at three typical values:
the central and the two extremes within the 1-$\sigma$ region of each fit.
For convenience, we define a function of $g_3^\prime$,
 $g(g_3^\prime)\equiv1.25+24|g_3^\prime|(M_D+M_\pi)M_\pi/\bar{M}_D$, and rewrite
 the bounds in Eq.~(\ref{Boundp3}) as
\bea
\left\{
\begin{array}{r}
h_4^\prime-h_5^\prime\geq g(g_3^\prime)\ ,\\
h_4^\prime\geq0.55\ .
\end{array}
\right.\label{Boundp3forplot}
\eea
Notice that the bounds depend on the renormalization scale $\mu$ since the loop
contributions $f_{\rm process}^{(2)}(s,t)$ are involved. The NNLO bounds in
Eq.~(\ref{Boundp3}) and Eq.~(\ref{Boundp3forplot}) are obtained by setting
$\mu=\bar{M}_D$ in accordance with Ref.~\cite{Yao:2015qia}.
The comparison is shown in Fig.~\ref{fig:boundP3}.
The bounds displayed in the graphs in the first,
second and third column correspond to taking the central, the lowest and
the largest value in each fit for $g_3^\prime$, respectively. As seen from the
plots, no fit completely obeys the bounds. For UChPT-6($a$)  and UChPT-6($b$),
the fit values are consistent with the first bound in Eq.~(\ref{Boundp3})
while they violate the second one, i.e., the one restricting $h_4'$ only.
Both bounds are violated in the fits for UChPT-6($a^\prime$)  and
UChPT-6($b^\prime$), which are the ones with a prior, which requires all of the
LECs (made dimensionless) to be take natural values of order $\order{1}$.

These comparisons, however, have to be interpreted with caution.
The positivity bounds in Eq.~\eqref{PosboundsNLO} and~\eqref{Boundp3} were
derived using the perturbative scattering amplitudes, while the fits in
Ref.~\cite{Liu:2012zya,Altenbuchinger:2013vwa,Yao:2015qia} were performed using
resummed amplitudes with perturbative kernels. The resummed amplitudes using
various unitarization approaches in the literature break the crossing symmetry,
which, however, is one of the main components in deriving the positivity bounds
through dispersion relations. It is thus not surprising that the LECs determined
in the UChPT fits do not respect the positivity bounds. Nevertheless, we notice
that all of these fits prefer a negative value for $h_4$ while the positivity
bound requires it to be positive.

\section{Summary and outlook \label{SecCon}}

We have estimated the LECs in the NLO and NNLO chiral Lagrangian for the $D\phi$
interaction using resonance exchanges. These LECs receive contributions from
exchanging the scalar charmed mesons, the light-flavor vector, scalar and tensor
mesons. We found that $h_1$ is entirely saturated by the light scalar-meson
exchange. The resulting estimates are consistent with the NLO UChPT
fitting results~\cite{Liu:2012zya, Altenbuchinger:2013vwa}, while sizeable
deviations from the determinations with the NNLO UChPT~\cite{Yao:2015qia} are
found. 
More lattice data on the $D\phi$ scattering observables would be useful to
better pin down the LECs at NNLO.

In parallel, with the help of axiomatic $S$-matrix principles, such as
unitarity, analyticity and crossing symmetry, we derived positivity constraints
on the $D$$\pi$ and $D_s$$\pi$ scattering amplitudes in upper parts of
Mandelstam triangles, $\mathcal{R}_{D\pi}$ and $\mathcal{R}_{D_s\pi}$,
respectively. In combination with the corresponding scattering amplitudes
calculated in ChPT using the EOMS scheme, the constraints are then translated
into a set of bounds on the LECs.
At order $\mathcal{O}(p^2)$, the bounds are independent of the Mandelstam
variables $s,t$ and hence have unique forms throughout $\mathcal{R}_{D\pi}$ or
$\mathcal{R}_{D_s\pi}$. At order $\mathcal{O}(p^3)$, the most stringent bounds
are obtained by zooming inside the upper part of the Mandelstam triangle such
that they can easily be employed and implemented to constrain future analyses.
Finally, as a first use of these bounds, the values of LECs in the literature
are compared with them. The comparison shows that the bounds, in particular the
one constraining $h_4$ only, are badly violated in all the previous
determinations from fitting to lattice data using UChPT.
The most probable reason for this is that the UChPT amplitudes violate
crossing symmetry which is the basis of deriving the positivity bounds.

For a more reasonable comparison, one needs to derive positivity bounds for the
unitarized amplitudes. One possible attack to the problem could come from using
the method proposed in Ref.~\cite{Hannah:2001ee} where the author proposed a
crossing-symmetric amplitude for the process $\gamma\pi\to \pi\pi$ combining the
inverse amplitude method, which is one of the unitarization approaches, and the
Roy equation.
In our case, the problem is much more involved due to different masses and
coupled channels. Whether such a method can lead to a feasible procedure still
needs to be seen.

\section*{Acknowledgements}
We would like to thank Z.-H.~Guo and W.~Chen for helpful discussions.
This work is supported in part by DFG and NSFC through funds provided to the
Sino-German CRC 110 ``Symmetries and the Emergence of Structure in QCD" (NSFC
Grant No.~11621131001), by the Thousand Talents Plan for Young Professionals, by
the CAS Key Research Program of Frontier Sciences (Grant No.~QYZDB-SSW-SYS013),
and by the CAS President's International Fellowship Initiative (PIFI) (Grant
No.~2015VMA076).

\bigskip

\appendix
\section{LO Born Amplitudes\label{app:LOBorn}}

In this appendix, the LO Born amplitudes for various resonance exchanges
are listed for completeness. For a given amplitude, we use capital
subscripts, $S$, $T$ and $U$, to label the channels and superscripts,
$D_0^\ast$, $V$ (vector) and $S$ (scalar), to mark which resonance is exchanged.
The coefficients appearing in the amplitudes are listed in
Table~\ref{tab:AmpResCoes}.

\begin{table*}[htbp]
\caption{Coefficients for the resonance-exchange
amplitudes.}\label{tab:AmpResCoes}
\begin{center}
\newcolumntype{C}{>{\centering\arraybackslash}X}
\newcolumntype{R}{>{\raggedleft\arraybackslash}X}
\renewcommand{\arraystretch}{1.3}
\begin{tabularx}{\textwidth}{lCCCcCCC}
\hline\hline
Physical Processes		&$\mathcal{C}_S^{D_0^\ast}$ & $\mathcal{C}_U^{D_0^\ast}$	&$\mathcal{C}_T^V$	&$\mathcal{C}_{T,m}^{S8}$ & $\mathcal{C}_{T,d}^{S8}$& $\mathcal{C}_{T,m}^{S1}$ & $\mathcal{C}_{T,d}^{S1}$\\
\hline
$D^0K^-\to D^0K^-$		&$0$				&$2$				&$\sqrt{2}$	&$M_K^2$					&$1$		&$M_K^2$ & $1$\\
$D^+K^+\to D^+K^+$	&$0$				&$0$				&$0$			&$-2M_K^2$					&$-2$	&$M_K^2$ &$1$\\
$D^+\pi^+\to D^+\pi^+$	&$0$				&$2$				&$\sqrt{2}$	&$M_\pi^2$					&$1$		&$M_\pi^2$ &$1$\\
$D^+\eta\to D^+\eta$	&$\frac{1}{3}$		&$\frac{1}{3}$		&$0$			&$\frac{1}{3}(5M_\pi^2-8M_K^2)$	&$-1$	&$M_\eta^2$ &$1$\\
$D_s^+K^+\to D_s^+K^+$	&$0$				&$2$				&$\sqrt{2}$	&$M_K^2$					&$1$		&$M_K^2$ &$1$\\
$D_s^+\eta\to D_s^+\eta$	&$\frac{4}{3}$		&$\frac{4}{3}$		&$0$			&$\frac{2}{3}(8M_K^2-5M_\pi^2)$	&$2$		&$M_\eta^2$ &$1$\\
$D_s^+\pi\to D_s^+\pi$	&$0$				&$0$				&$0$			&$-2M_\pi^2$					&$-2$	&$M_\pi^2$ &$1$\\
$D^0\eta\to D^0\pi^0$	&$\sqrt{\frac{1}{3}}$	&$\sqrt{\frac{1}{3}}$	&$0$			&$\sqrt{3}M_\pi^2$				&$\sqrt{3}$&$0$ &$0$\\
$D_s^+K^-\to D^0\pi^0$	&$\sqrt{2}$		&$0$				&$-1$		&$\frac{3}{2\sqrt{2}}(M_K^2+M_\pi^2)$&$\frac{3}{\sqrt{2}}$&$0$ &$0$\\
$D_s^+K^-\to D^0\eta$	&$\sqrt{\frac{2}{3}}$	&$-2\sqrt{\frac{2}{3}}$&$-\sqrt{3}$	&$\sqrt{\frac{3}{8}}(3M_\pi^2-5M_K^2)$&$-\sqrt{\frac{3}{2}}$&$0$ &$0$\\
\hline\hline
\end{tabularx}
\end{center}
\end{table*}

\begin{itemize}
\item{$D_0^\ast$ exchange: ($m=\overset{\circ\;\;\;}{M_{D_0^\ast}}$)}
\bea
\mathcal{A}_S^{D_0^\ast}(s,t,u)&=&\mathcal{C}_S^{D_0^\ast}\frac{{g}_0^2}{F_0^2}\frac{p_1\cdot p_2\,p_3\cdot p_4}{s-m^2}\ ,\nonumber\\
\mathcal{A}_U^{D_0^\ast}(s,t,u)&=&\mathcal{C}_U^{D_0^\ast}\frac{{g}_0^2}{F_0^2}\frac{p_1\cdot p_4\,p_2\cdot p_3}{u-m^2}\ .
\eea
\item{Light-flavor vector meson exchange:}
\bea
\mathcal{A}_T^{V}(s,t,u)&=&C_T^V\frac{g_{DDV}g_V}{F_0^2}\left\{\frac{(p_1-p_3)\cdot p_2\,(p_1+p_3)\cdot p_4}{M_V^2-t}-(p_2\leftrightarrow p_4)\right\}\ .
\eea
\item{Light-flavor scalar meson exchange:}
\bea
\mathcal{A}_T^{S}(s,t,u)&=&\frac{2{g}_{DDS}\left\{\mathcal{C}_{T,m}^{S8}-\mathcal{C}_{T,d}^{S8}\, p_2\cdot p_4\right\}}{3F_0^2(M_{S8}^2-t)}+\frac{4\tilde{g}_{DDS}\left\{\mathcal{C}_{T,m}^{S1}-\mathcal{C}_{T,d}^{S1}\, p_2\cdot p_4\right\}}{F_0^2(M_{S1}^2-t)}\ .
\eea
\end{itemize}

\section{Estimate of the tensor resonance contribution to the LEC $h_5$
\label{app:h5}}

In this appendix, we use the technique in Ref.~\cite{Ecker:1988te,Ecker:1989yg},
which is different from but equivalent to the one used in the main text, to
estimate the contribution of exchanging the light tensor mesons, denoted by $T$,
with $J^{PC}=2^{++}$ to the LEC $h_5$. We construct the Lagrangian for the $DDT$
coupling as
\begin{equation}
\mathcal{L}_{DDT}= g_{DDT} \mathcal{D}_{\mu} D T^{\mu\nu} \mathcal{D}_\nu
D^\dagger. \label{lddt}
\end{equation}
One could calculate the tensor-meson contribution to the LECs by integrating
out the tensor meson field as follows.

The $J^{PC}=2^{++}$ mesons are described by the symmetric hermitian
field~\cite{Ecker:2007us}:
\begin{equation}
T_{\mu\nu}=T^0_{\mu\nu}\frac{\lambda_0}{\sqrt{2}}+\frac{1}{\sqrt{2}}
\sum^8_{i=1}\lambda_iT^i_{\mu\nu}, ~~T_{\mu\nu}=T_{\nu\mu}\,,
\end{equation}
where the singlet and octet components are
\begin{equation}
T^0=f^0_2, \quad \text{and}\quad \frac{1}{\sqrt{2}}\sum^8_{i=1}\lambda_i
T^i=\begin{pmatrix} \frac{a^0_2}{\sqrt{2}}+\frac{f_2^8}{\sqrt{6}} & a_2^+ &
K_2^{*+} \\
	a_2^- & -\frac{a_2^0}{\sqrt{2}}+\frac{f_2^8}{\sqrt{6}} & K_2^{*0} \\
	K_2^{*-} & \bar{K}_2^{*0} & -\frac{2f_2^8}{\sqrt{6}}
\end{pmatrix},
\end{equation}
respectively.

The coupling of a single tensor meson to the Goldstone bosons can be described
by the following Lagrangian~\cite{Ecker:2007us}
\begin{equation}\label{eq:LagpipiT}
\mathcal{L}=-\frac{1}{2}\langle T_{\mu\nu} D_T^{\mu\nu,\rho\sigma}T_{\rho\sigma}\rangle
+\langle T_{\mu\nu}J^{\mu\nu}_T\rangle,\qquad J_T^{\mu\nu}\equiv g_T \{ u^\mu ,u^\nu \}\ ,
\end{equation}
where $J_T^{\mu\nu}$ is the tensor current  and
\begin{equation}
\begin{split}
D_T^{\mu\nu,\rho\sigma}\,=\,&(\mathcal{D}^2+M_T^2)\left[
\frac{1}{2}(g^{\mu\rho}g^{\nu\sigma}+g^{\mu\sigma}g^{\nu\rho})-g^{\mu\nu}
g^{\rho\sigma}\right] \\
&+g^{\rho\sigma}\mathcal{D}^\mu \mathcal{D}^\nu+g^{\mu\nu}\mathcal{D}^\rho \mathcal{D}^\sigma
-\frac{1}{2}(g^{\nu\sigma}\mathcal{D}^\mu \mathcal{D}^\rho +g^{\rho\nu}\mathcal{D}^\mu \mathcal{D}^\sigma+g^{\mu\sigma}\mathcal{D}^\rho \mathcal{D}^\nu+g^{\rho\mu}\mathcal{D}^\sigma \mathcal{D}^\nu).
\end{split}
\end{equation}
Inserting the equation of motion for the tensor mesons,
$T_{\rho\sigma}=(D^{\mu\nu,\rho\sigma})^{-1}J_T^{\mu\nu}$, into
$\mathcal{L}_{DDT}$, we get
\begin{equation}
\begin{split}
&\mathcal{L}_{DDT}=\mathcal{L}_{DDT}^{(2)}+O(p^4), \\
&\mathcal{L}_{DDT}^{(2)}=\frac{g_{DDT}}{M_T^2} \mathcal{D}_{\mu} D J_T^{\mu\nu} \mathcal{D}_\nu D^\dagger
=\frac{g_{DDT} g_T}{M_T^2} \mathcal{D}_{\mu} D \{ u^\mu ,u^\nu \} \mathcal{D}_\nu D^\dagger.
\end{split}\label{eq:tensorlag100}
\end{equation}
It is then easy to see that the light-tensor mesons only contribute to the LEC
$h_5$, and the contribution is
\bea h_5^T=\frac{ g_{DDT} g_T}{M_T^2}\, .\eea

\section{Estimate of the coupling constant {\boldmath$g_{DDT}$} via QCD sum rules\label{app:sumrule}}

In this Appendix, we estimate the unknown off-shell coupling constant $g_{DDT}$
using QCD sum rules, following the procedure in, e.g.,
Refs.~\cite{Bracco:2011pg, Azizi:2014yua}. %
To be specific, we will calculate the $D^0 D^-
a_2^+$ coupling. The standard procedure for computing a coupling constant in the
method of QCD sum rules is to consider the three-point correlation
function, which in our case is given by
\begin{equation}
\Pi_{\mu \nu}(p',p,q)=i^2 \int d^4 x \int d^4y\, e^{i(-p' x+yp)}
\langle0|T\{j^{D^0}(x)j^{D^-}(y) j^{a^+_2}_{\mu \nu}(0) \}|0\rangle,
\label{equpi}
\end{equation}
where $q=p'-p$ denotes the momentum transfer. The interpolating currents
that we use for the $D^0$, $D^-$ and $a_2^+$ mesons are
\begin{equation}
\begin{split}
&j^{D^0}(x)=i \bar{u}(x) \gamma_5 c(x)\,,\\
&j^{D^-}(x)=i \bar{c}(x) \gamma_5 d(x)\,,\\
&j^{a_2^+}_{\mu\nu}(x)=\frac{i}{2} \bar{d}(x) (\gamma_\mu
\stackrel{\leftrightarrow}{D}_\nu +\gamma_\nu \stackrel{\leftrightarrow}{D}_\mu
)u(x)\,,
\end{split}
\end{equation}
where $\stackrel{\leftrightarrow}{D}_\mu=({\overrightarrow
D}_\mu-{\overleftarrow D}_\mu)/2$.

One can calculate the correlation function in two different ways. On the one
hand, the correlation function in Eq.~\eqref{equpi} can be computed by inserting a
complete sets of appropriate hadronic states with the same quantum numbers as
the interpolating currents. Following the usual procedure, we obtain
\begin{equation}
\begin{split}
\Pi^{had}_{\mu\nu}(p'^2,p^2,q^2)=& \frac{\langle0|j^{D^0}|D^0(p')\rangle
\langle0|j^{D^-}|D^-(p)\rangle \langle 0|
j_{\mu\nu}^{a_2^+}|a_2^+(q,\epsilon)
\rangle}{(p'^2-m_D^2)(p^2-m_D^2)(q^2-m_a^2)}\\
& \times \langle D^0(p') a_2^+(q,\epsilon)|D^+(p)\rangle +
\ldots\,,\label{hadsideone}
\end{split}
\end{equation}
where the ellipses represent the contributions of the excited states and the
continuum.
The matrix elements above are parameterized as follows~\cite{Azizi:2014yua}
\begin{equation}
\begin{split}
&\langle 0|j^{D}|D(p)\rangle=i\frac{m_D^2 f_D}{m_c+m_q},\\
&\langle 0 |j_{\mu\nu}^{a_2^+}|a_2^+(q,\epsilon) \rangle=m^3_a f_a
\epsilon^{*(\lambda)}_{\mu\nu},\\
&\langle D^0(p')
a_2^+(q,\epsilon)|D^+(p)\rangle=g_{DDa_2}\epsilon^{(\lambda)}_{\alpha\beta}
p'_\alpha p_\beta,
\end{split}
\end{equation}
where $f_D$, $f_a$ are the decay constants of $D^0(D^-)$ and $a_2^+$ mesons, and
$g_{DDa_2}$ is the form factor of the $DDT$ coupling under consideration.
Substituting the above matrix elements into Eq.~\eqref{hadsideone}, the
correlation function takes the form
\begin{equation}
\begin{split}
\Pi^{had}_{\mu\nu}(p'^2,p^2,q^2)&= i^2\left(\frac{m_D^2
f_D}{m_c+m_q}\right)^2\frac{g_{DDa_2} f_a
m_a^3}{(p'^2-m_D^2)(p^2-m_D^2)(q^2-m_a^2)}
\\
& \times \left\{1-\frac{1}{3 m_a^2}(p^2+p'^2+2 q^2)-\frac{1}{3 m_a^4}\left[
(p^2-p'^2)^2-(q^2)^2\right] \right\}( p'^\mu p^\nu+p'^\nu
p^\mu)+ \ldots \, , \label{hadronicside}
\end{split}
\end{equation}
where only the Lorentz structure $( p'^\mu p^\nu+p'^\nu p^\mu)$ is kept, and
the following relation has been used,
\begin{equation}
\sum_{\lambda}
\epsilon_{\mu\nu}^{(\lambda)}\epsilon^{*(\lambda)}_{\alpha\beta}=\frac{1}{2}T_{
\mu\alpha}T_{\nu\beta}+
\frac{1}{2}T_{\mu\beta}T_{\nu\alpha}-\frac{1}{3}T_{\mu\nu}T_{\alpha\beta},
\end{equation}
with $T_{\mu\nu}=-g_{\mu\nu}+q_{\mu}q_{\nu}/m_a^2$.

On the other hand, the
correlation function can be calculated at the quark-gluon
level using the QCD operator product expansion (OPE) method. It is convenient to
evaluate it in the fixed-point gauge: $(x-x_0)^\mu A_\mu^a(x)=0$, where $x_0$
is an arbitrary point in the coordinate space and could be chosen at the origin.
Then, in the deep Euclidean region, the potential can be expressed in
terms of the field strength tensor $G_{\mu \nu}={\lambda^a}G^a_{\mu
\nu}/2$ as~\cite{Reinders:1984sr}
\begin{equation}
A_\mu (x)=\frac{1}{2}x^\nu G_{\nu\mu}(0)+\frac{1}{3}x^\alpha x^\nu D_\alpha
G_{\nu\mu}(0)+ \order{x^3}\, .
\end{equation}

Since we are not aiming at a precise calculation, we will only keep the vacuum
condensate of the lowest dimension, that is the quark condensate.
Considering only the Lorentz structure $(p'^\mu p^\nu+p'^\nu p^\mu)$ and using
the double dispersion relation,  we find
\begin{equation}
\Pi_{\mu\nu}(p'^2,p^2,q^2)=\Pi(p'^2,p^2,q^2)(p'^\mu p^\nu+p'^\nu p^\mu)+ ...,
\end{equation}
\begin{equation}
\Pi(p'^2,p^2,q^2)=\int ds_1 ds_2
\frac{\rho^\text{pert}(s_1,s_2,q^2)}{(s_1-p'^2)(s_2-p^2)}+\Pi^{qq}(p'^2,p^2,q^2),
\label{PiOPE}
\end{equation}
where
\begin{equation}
\begin{split}
\rho^\text{pert}(s_1,s_2,q^2)=& -\frac{3}{8 \pi^2 \lambda^{5/2}}(s_1+s_2-t-2
m_c^2)\Big\{ (s_1 s_2+m_c^4) (\lambda +3t(s_1+s_2-t)) \\
&-3m_c^2(s_1+s_2-t)\left[(s_1-s_2)^2-t(s_1+s_2)\right] \Big\} ,
\end{split}
\end{equation}
with $ \lambda=(s_1+s_2-t)^2-4 s_1 s_2 $, and
\begin{equation}
\Pi^{qq}(p'^2,p^2,q^2)=\frac{1}{4} m_c\langle \bar{q}q \rangle
\left(\frac{1}{p'^2-m_c^2}+\frac{1}{p^2-m_c^2}
\right)\frac{1}{q^2}\, .
\end{equation}

In order to suppress the  contribution from the excited states, we perform a
double Borel transformation in both variables $p'^2$ and $p^2$ to
the correlation functions in Eqs.~\eqref{hadronicside} and \eqref{PiOPE}.
Using the quark-hadron duality, we obtain
\begin{eqnarray}
\Pi(M_B^2,M'^2_B,q^2)\al=\al i^2\,
\frac{g_{DDa_2} f_a m_a^3}{q^2-m_a^2}\left(\frac{m_D^2 f_D}{m_c+m_q}\right)^2
\left[
1-\frac{2(q^2+m_D^2)}{3m_a^2}+\frac{q^4}{3m_a^4}\right] e^{ -m_D^2/M_B^2
-m_D^2/M'^2_B} \nonumber\\
\al=\al \int^{s_1^0}_{s_{1\text{min}}} \int^{s_2^0}_{s_{2\text{min}}}ds_1 ds_2
\rho^\text{pert}(s_1,s_2,q^2)e^{-s_1/M_B^2-s_2/M'^2_B}\, .
\end{eqnarray}

It is clear that the coupling $g_{DDa_2}$ is in fact given by  a form factor as
a function of the Euclidean momentum $Q^2=-q^2$ which will be denoted by
$g_{DDa_2}^{(a_2)}(Q^2)$, where the superscript means
that the meson $a^+_2$ is off-shell while the $D$ mesons are on-shell since the
correlation function is evaluated in the space-like region $Q^2>0$.

We neglect the light quark masses and use the following values for numerical
analysis: $m_c=1.27~\mbox{GeV}$, $m_D=1.87~\mbox{GeV}$,  $m_a=1.32~\mbox{GeV}$,
$\langle \bar{q}q\rangle=(-0.24)^3~\mbox{GeV}$,
$f_D=0.207~\mbox{GeV}$~\cite{Azizi:2014yua}, and
$f_a=0.041$~\cite{Cheng:2010hn}. Furthermore, $s_{1\text{min}}=m_c^2$ and
$s_{2\text{min}}=\frac{m_c^2}{m_c^2-s_1}q^2+m_c^2$.
Since the dependence of the form factor on $M_B^2$ and $M'^2_B$ is weak, one can
use set $M'^2_B=M_B^2$~\cite{Bracco:2011pg}.

The window for the Borel mass $M_B^2$ can be determined by requiring both the
dominance of the ground state hadronic poles and the convergence of OPE. The
quark condensate contribution would disappear if the double Borel transformation
is performed in the variables $p'^2$ and $p^2$. In order to estimate the lower
bound of $M_B^2$, we choose to perform the double Borel transformation in
variables $p^2$ and $q^2$, and assume the lower bound is same  as that in the
double Borel transformation in $p'^2$ and $p^2$.
The lower limit of $M_B^2$ is estimated by requiring
$|\Pi^{qq}(p'^2,M_B^2,M_B^2)/\Pi^\text{pert}(p'^2,M_B^2,M^2_B)|$ to be smaller
than 25\% for Euclidean momentum $p'^2$.
 At the same time, the upper bound of the Borel mass $M_B^2$ can be estimated
by requiring the pole contribution (PC) to be larger than 75\% which is
defined by
\begin{equation}
\text{PC}=\frac{\int^{s_1^0}_{s_{1\text{min}}}ds_1\int^{s_2^0}_{s_{2\text{min}}}ds_2
\rho^\text{pert}(s_1,s_2,q^2)e^{-s_1/M_B^2-s_2/M_B^2}}{\int^{\infty}_{s_{1\text{min}}}ds_1
\int^{\infty}_{s_{2\text{min}}}ds_2
\rho^\text{pert}(s_1,s_2,q^2)e^{-s_1/M_B^2-s_2/M_B^2}}\,.
\end{equation}

\begin{figure}
\begin{center}
\includegraphics[width=0.6\linewidth]{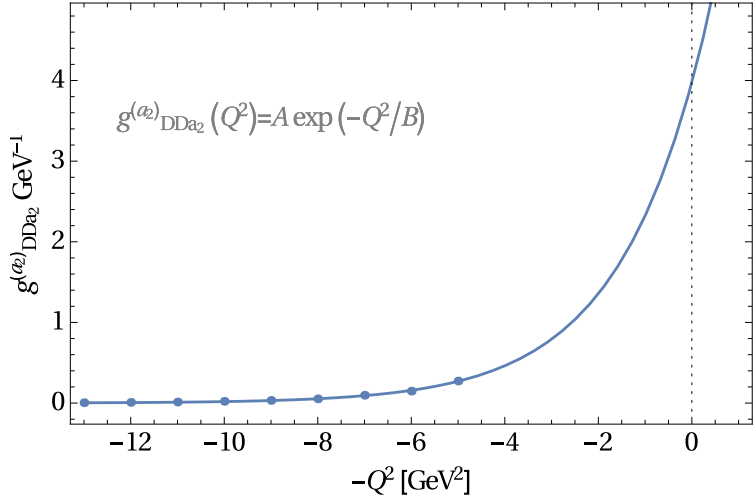}
\caption{Momentum dependence of the $DDa_2$ form factor (for off-shell $a_2$).
The dots give the results from QCD sum rules, and the solid line gives the
extrapolation.}
\end{center}
\end{figure}

The parameters $s_1^0$ and $s_2^0$ are chosen around the region
where the variation of coupling constant $g_{DDa_2}^{(a_2)}(Q^2)$ is minimal.
Given a value of $Q^2$, we obtain a corresponding $g_{DDa_2}^{(a_2)}(Q^2)$. From
the above requirements, the Borel window we use here is $M^2_B\sim [
3.2~\mbox{GeV}^2, 4.0~\mbox{GeV}^2 ] $. We take $M^2_B=
 3.6~\mbox{GeV}^2$ for estimating the form factor $g_{DDa_2}^{(a_2)}(Q^2)$,
 and the values of $s_1^0$ and $s_2^0$ are chosen to increase slightly from
 around $6.0~\mbox{GeV}^2$ to $8.5~\mbox{GeV}^2$ as increasing $Q^2$ from
 $5~\mbox{GeV}^2$ to $12~\mbox{GeV}^2$.  Since we
are only able to calculate the form factor in the deep Euclidean region,
 we need to extrapolate it to $Q^2=0$
to get the coupling constant. The extrapolation is rather model-dependent. To be
specific, we simply take the form
$g_{DDa_2}^{(a_2)}(Q^2)= A \exp(-{Q^2}/{B})$ used in
Refs.~\cite{Bracco:2011pg, Yu:2015xwa} despite that no physical reasoning is
behind this parametrization.
With this form, we fit to a few points in the Euclidean region, and get
$A=10.1~\mbox{GeV}^{-1}$ and $B=1.9~\mbox{GeV}^2$. Finally, we get an estimate
for the coupling constant as
\begin{equation}
 g_{DDT}\approx g_{DDa_2}(0)\approx 3.9~\mbox{GeV}^{-1}.
\end{equation}
It should be noted that such an estimate bears a large uncertainty which we do
not know how to quantify, and the resulting value can only be regarded as an
order-of-magnitude estimate.


\end{document}